\documentclass[prx,twocolumn,nofootinbib,citeautoscript,longbibliography,notitlepage,superscriptaddress]{revtex4-1}

\usepackage{graphicx}
\usepackage{dcolumn}
\usepackage{bm}
\usepackage{color}
\usepackage{amsmath}
\usepackage{tabularx}
\usepackage{epstopdf}
\usepackage{latexsym}
\usepackage{amssymb}
\usepackage{amsmath}
\usepackage{colortbl}
\usepackage{psfrag}
\usepackage{bbm}
\usepackage{bm}
\usepackage{titlesec}
\usepackage{dsfont}
\usepackage{feynmp}
\usepackage{slashed}
\usepackage{multirow}
\usepackage[tight]{subfigure}

\usepackage{comment}

\usepackage[papersize={8.5in,11in}]{geometry}

\usepackage{color}
\definecolor{darkblue}{rgb}{0.,0.,0.4}
\definecolor{darkred}{rgb}{0.5,0.,0.}
\definecolor{BlueViolet}{RGB}{138,43,226}
\definecolor{SkyBlue}{RGB}{30,144,255}
\definecolor{DarkGreen}{RGB}{0,100,0}
\usepackage[pdftex,colorlinks=true,linkcolor=darkblue,citecolor=blue,urlcolor=darkred]{hyperref}

\usepackage{physics}

\renewcommand{\epsilon}{\varepsilon}

\allowdisplaybreaks[3] 

\usepackage{CJK}

\usepackage{amsmath}
\usepackage{feynmp}
\usepackage{amssymb}
\usepackage{epsfig}
\usepackage{graphicx}
\usepackage{subfigure}
\usepackage{mathrsfs}
\usepackage{hyperref}
\usepackage{cleveref}
\usepackage{cases}
\usepackage{bbm}
\usepackage{xcolor}
\usepackage{tabularx}
\usepackage{array}
\usepackage{booktabs}
\usepackage{multirow}
\usepackage{lipsum}
\usepackage{float}
\usepackage{lmodern}
\usepackage{textcomp}
\usepackage{makecell}
\usepackage{comment}
\usepackage{threeparttable}

\allowdisplaybreaks[4]

\begin{document}

\begin{CJK*}{GBK}{song}


\title{Low-energy critical behavior in two-dimensional tilted semi-Dirac semimetals driven by fermion-fermion interactions}

\date{\today}

\author{Wen Liu}
\altaffiliation{These authors contributed equally to this work.}
\affiliation{Department of Physics, Longyan University, Longyan, 364012, P.R. China}
\affiliation{Department of Physics, Tianjin University, Tianjian 300072, P.R. China}

\author{Wen-Hao Bian}
\altaffiliation{These authors contributed equally to this work.}
\affiliation{School of Physics, Nanjing University, Nanjing, Jiangsu 210093, P.R. China}

\author{Xiao-Zhuo Chu}
\altaffiliation{These authors contributed equally to this work.}
\affiliation{Department of Physics, Tianjin University, Tianjian 300072, P.R. China}

\author{Jing Wang}
\altaffiliation{Corresponding author: jing$\textunderscore$wang@tju.edu.cn}
\affiliation{Department of Physics, Tianjin University, Tianjian 300072, P.R. China}
\affiliation{Tianjin Key Laboratory of Low Dimensional Materials Physics and Preparing Technology,
Tianjin University, Tianjin 300072, P.R. China}

\begin{abstract}

Employing the renormalization group approach, we carefully investigate the critical
behavior of two-dimensional tilted semi-Dirac semimetals induced by the fermion-fermion interactions
in the low-energy regime. After incorporating all one-loop corrections, we derive the coupled RG
equations of all related parameters and introduce two distinct strategies, named as Strategy I and
Strategy II, to describe different scenarios. A detailed numerical analysis yields several interesting
behavior in the low-energy limit. At first, we notice that the fermion-fermion interactions
either vanish or diverge in the Strategy I, depending on the initial values of the tilting parameter and
the fermionic couplings, whereas these interactions in the Strategy II always diverge at a certain critical energy scale, which
is associated with the initial conditions. Next, the microstructural parameter $\alpha$ and
the fermion velocity $v_F$ in the Strategy I share the similar behavior with their Strategy II counterparts. It is observed that fermion-fermion interactions lead to an increase in $\alpha$ while driving a decrease in $v_F$.
Furthermore, the system can either be attracted by the Gaussian fixed point (GFP)
or certain relatively fixed point (RFP) in the Strategy I. However, it always flows toward
the RFP in the Strategy II at the lowest-energy limit.
These results would provide helpful insights into the studies on observable quantities and phase transitions
in the two-dimensional tilted semi-Dirac semimetals and the analogous semimetals.


\end{abstract}


\maketitle


\section{Introduction}

In recent years, extensive research including both theoretical and experimental studies has been devoted to
investigating Dirac materials~\cite{Novoselov2004Science,Novoselov2005Nature,Neto2009RMP,Burkov2011PRL,
Neto2009RMP,Peres2010RMP,Burkov2011PRL,Yang2011PRB,Wan2011PRB,Huang2015PRX,Hasan2015Science,
Xu2015Nature,Lv2015NP,Weng2015PRX,Roy2018PRX,Vafek2014ARCMP,Wehling2014AP,
Wang2012PRB,Young2012PRL,Steinberg2014PRL,Liu2014NM,Liu2014Science,Xiong2015Science,Roy2009PRB,
Roy2016,Roy-2014-2016,Savary2014PRB,Moon2014PRX,Montambaux,Vafek2014ARCMP,Hasegawa2006PRB}. These materials are characterized by the
presence of several discrete Dirac nodal points, such as two-dimensional (2D) graphene~\cite{Novoselov2004Science,Novoselov2005Nature,
Neto2009RMP,Peres2010RMP}, Weyl semimetals (WSMs)~\cite{Burkov2011PRL,Yang2011PRB,Wan2011PRB,Huang2015PRX,Hasan2015Science,
Xu2015Nature,Lv2015NP,Weng2015PRX,Roy2018PRX}, and Dirac semimetals~\cite{Vafek2014ARCMP,Wehling2014AP,
Wang2012PRB,Young2012PRL,Steinberg2014PRL,Liu2014NM,Liu2014Science,Xiong2015Science,Roy2009PRB,
Roy2016,Roy-2014-2016,Savary2014PRB,Moon2014PRX,Montambaux,Hasegawa2006PRB}.
A newly identified variant of Dirac materials, denominated as 2D semi-Dirac semimetals (SDSMs), has garnered particular attention~\cite{Hasegawa2006PRB,Pardo2009PRL,Katayama2006JPSJ,Dietl2008PRL,Delplace2010PRB,Wu2014Expre,
Banerjee2009PRL,Saha2016PRB,Uchoa1704.08780,Katayama2006JPSJ},
which possess the 2D Dirac-like properties with the gapless excitations and an anisotropic dispersion. Specifically,
the dispersion is parabolic in one direction and linear in the other. This unique behavior has been observed in various systems, including the quasi-two-dimensional organic conductor $\alpha-(\mathrm{BEDT-TTF)_2I_3}$ salt under uniaxial pressure~\cite{Katayama2006JPSJ}, tight-binding honeycomb lattices in the presence of a magnetic field~\cite{Dietl2008PRL}, $\mathrm{VO_2-TiO_2}$ multilayer systems (nanoheterostructures)~\cite{Pardo2009PRL}, and photonic systems consisting of a square array of elliptical dielectric cylinders~\cite{Wu2014Expre}. Besides, it is of remarkable interest to highlight that the Dirac/Weyl cones inherent in these materials
can be stretched and then tilted by breaking the certain symmetry with additional
forces~\cite{Lee2018PRB,Lee2019PRB,Jafari2019PRB-t}. This gives rise to the tilted Dirac semimetals, including mechanically deformed
graphene~\cite{Katayama2006JPSJ,Kobayashi2007JPSJ},
tilted $\mathrm{WTe}_{2}$~\cite{Soluyanov2015Nature}, the Fulde-Ferrell ground state of a spin-orbit coupled fermionic superfluid~\cite{Xu2015PRL}, or a cold-atom optical lattice~\cite{Xu2016PRA}.
Recently, these tilted Dirac materials have attracted considerable attention~\cite{Shekhar2015NP,Parameswaran2014PRX,Potter2014NC,Baum2015PRX,
Arnold2016NC,Zhang2016NC,Lee2018PRB,Lee2019PRB,Fritz2017PRB,Fritz2019arXiv,
Jafari2018PRB,Alidoust2019arXiv,Yang2018PRB,Trescher2015PRB,Proskurin2015PRB,
Brien2016PRL,Zyuzin2016JETPL,Ferreiros2017PRB,Jafari2019PRB-t,Juricic2023JHEP}.
Attesting to the unique low-energy excitations exhibited by 2D SDSMs, one may anticipate that a number of
distinct behavior differs from those of the general Dirac fermions within the low-energy
regime~\cite{Neto2009RMP,Hasan2010RMP,Qi2011RMP,Huang2015PRX,Hasan2015Science,Ding2015NPhys}.

In principle, the 2D tilted SDSMs can be categorized into two scenarios heavily depending on the strength of tilting parameters~\cite{Neto2009RMP,Peres2010RMP,Jafari,Lee2018PRB,Jafari2019PRB-t,Lee2019PRB,Soluyanov2015Nature,Noh2017PRL,Fei2017PRB,Yan2017NC},
namely, type-I tilted SDSMs featured by intact Dirac points (cones)~\cite{Neto2009RMP,Peres2010RMP,Jafari} and
type-II tilted SDSM characterized by the open Fermi surface (such as two straight lines, etc.)
~\cite{Soluyanov2015Nature,Noh2017PRL,Fei2017PRB,Yan2017NC}.
Without loss of generality, we put our focus on the type-I situation within this work.
In this circumstance, it is of significance to investigate the critical behavior of such type-I
tilted SDSMs, which is caused by the unconventional low-energy excitations accompanied by
unique Dirac cones in tandem with their intimate interactions in the low-energy regime.
Regarding the interactions in Dirac materials, two types are particularly important: Coulomb
interactions and fermion-fermion interactions, which may play an important role in pinning down
their low-energy properties. In principle, the long-range Coulomb interaction can be easily screened in the realistic systems via
adding certain substrate to realize a large dielectric constant~\cite{Neto2009RMP,Kotov2012RMP,Sarma2011RMP,Katsnelson2006PRB}.
As a result, its effects on the low-energy properties of tilted SDSMs are more important    ~\cite{Lee2018PRB,Lee2019PRB,Fritz2017PRB,Fritz2019arXiv}. In comparison,
the fermion-fermion interactions can still be intact, and henceforth, 
they are closely associated with the unusual signatures in fermionic
systems~\cite{BXW2023EPJP,DZZW2020PRB,Kivelson2009PRL,Murray2014PRB,Herbut,Herbut-2,Wang2017PRB_QBCP,Wang2018,
Fradkin2010ARCMP,Roy2004arXiv,Nandkishore2012NP,Wang-NPhyB,Sur2016NJP,Roy2016JHEP,Han2017PRB,
Roy2018PRX2,Mandal2018PRB,Sur2019PRL,Szabo2021,Wang2023EPJB,Fu2023arXiv} while the Coulomb interaction is screened~\cite{Katsnelson2006PRB,Neto2009RMP,
Kotov2012RMP,Sarma2011RMP}. However, as to the 2D tilted SDSMs, it is worth highlighting that the fermion-fermion interactions
have hitherto been paid relatively less attention, which may lead to the neglect or omission of valuable physical information.
Accordingly, it is therefore of particular significance to systematically examine the role of
fermion-fermion interactions in determining the low-energy fates of 2D tilted SDSMs.
Unambiguously elucidating this issue would be greatly helpful to reveal
the unique properties of 2D semi-Dirac materials and even instructive to explore new
Dirac-like materials.

Inspired by the analysis presented above, we construct an effective field theory based on the microscopic model of
2D tilted SDSM by incorporating four distinct sorts of short-range fermion-fermion interactions
~\cite{Lee2018PRB,Wang2018,Montambaux,BXW2023EPJP,DZZW2020PRB,Kivelson2009PRL,Murray2014PRB,Herbut,Herbut-2,Wang2017PRB_QBCP,Wang2018,
Fradkin2010ARCMP,Roy2004arXiv,Nandkishore2012NP,Wang-NPhyB,Sur2016NJP,Roy2016JHEP,Han2017PRB,
Roy2018PRX2,Mandal2018PRB,Sur2019PRL,Szabo2021,Wang2023EPJB,Fu2023arXiv,Wang2023arXiv}.
In order to achieve an unbiased treatment for all the physical ingredients, we employ the
powerful energy-shell renormalization group (RG) approach~\cite{Wilson1975RMP,Polchinski1992arXiv,Shankar1994RMP}.
Following the standard procedures of the RG framework, we perform all one-loop calculations and derive the coupled RG evolutions of
all interaction parameters, which carry the full physical information in the low-energy regime. To simplify our study,
we designate a scaling parameter $\eta$ in Sec.~\ref{Sec_two_strategies}, which yields two distinct strategies
for the RG studies (Strategy I and Strategy II as defined in Sec.~\ref{Sec_two_strategies}).
Several interesting physical behavior have been obtained within these two strategies
in the low-energy regime.


At first, we find that the fermion-fermion interactions tend to either vanish or diverge in the low-energy regime.
In the Strategy I, the bigger initial values of the tilting parameter and the fermionic couplings cause fermion-fermion interactions to diverge at the low-energy limit. On the contrary, for smaller initial values, the fermion-fermion interactions diminish and eventually vanish as energy decreases. In comparison, as for the Strategy II, the fermion-fermion interactions always diverge at certain critical energy scale labeled by $l_c$, regardless of the initial parameter values. However, for both strategies,
the critical energy scale can be increased by the bigger initial values of fermionic interactions ($\lambda_{i0}$) and the tilting parameter $(\zeta)$, in tandem with the smaller values of the fermion velocity $(v_{F0})$.
Besides, a smaller microstructural parameter $(\alpha_0)$ increases the critical energy scale
in the Strategy I but instead decreases in the Strategy II.

Subsequently, we notice that the microstructural parameter $\alpha$ and the fermion velocity $v_F$ exhibit
the analogous behavior for both the Strategy I and Strategy II. It is observed that fermion-fermion interactions lead to an increase in $\alpha$ while driving a decrease in $v_F$. For the Strategy I, if the fermion-fermion interactions
do not diverge, $\alpha$ increases and $v_F$ gradually decreases in the low-energy regime.
In contrast, when the fermion-fermion interactions diverge
at the critical energy scale $l_c$ for either Strategy I or Strategy II, both $\alpha$ and $v_F$ are
required to stop at $l=l_c$ with taking values $\alpha(l_c)$ and $v_F(l_c)$, which are sensitively
influenced by the relevant parameters as collected in Table~\ref{Tab1} for various initial conditions.
Furthermore, the fixed point as addressed in Sec.~\ref{Subsec_RFP} is introduced
to capture the critical behavior of the system. We find that the 2D tilted SDSMs can either flow towards the Gaussian fixed point (GFP)
or certain relatively fixed point (RFP). With a fixed tilting parameter, the RFP is relatively robust against variations in other parameters. However, it is susceptible to the initial value of the tilting parameter $\zeta$ in both strategies.

The remainder of the paper is organized as follows. The effect model of the 2D tilted SDSM is
presented in Sec.~\ref{Sec_model}. We perform the energy-dependent RG analysis in Sec.~\ref{Sec_RGEqs}
and derive the coupled RG equations. In Sec.~\ref{Sec_two_strategies}, we introduce two strategies for simplifying
the analysis. Sec.~\ref{Sec_phyiscal_behavior} is followed to systematically examines the behavior
of all relevant parameters, including the fermion-fermion
interactions, microstructural parameter, and fermion velocity, and
then discuss the relatively fixed point (RFP) for both strategies.
Finally, we provide a brief summary in Sec.~\ref{Sec_summary}.

\begin{figure*}[htbp]
    \centering
    \subfigure[~$\zeta =0.1$]{\includegraphics[scale=0.5]{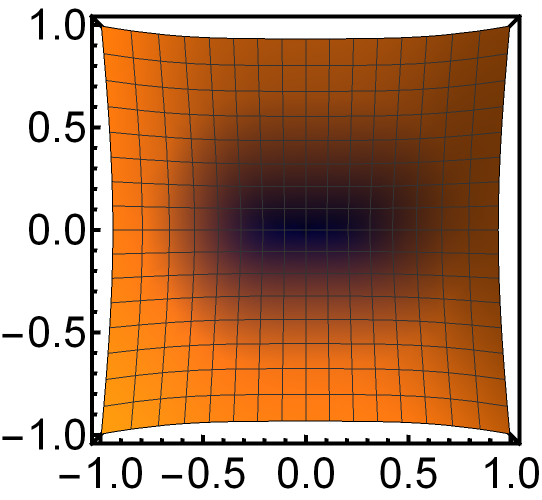}}
    \hspace{15pt}
    \subfigure[~$\zeta =0.5$]{\includegraphics[scale=0.5]{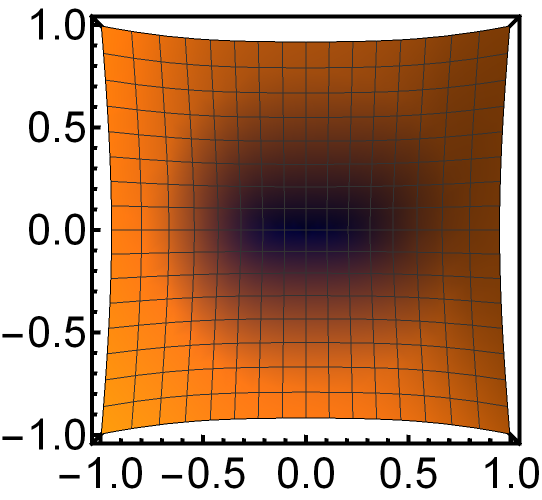}}
    \hspace{15pt}
    \subfigure[~$\zeta =0.9$]{\includegraphics[scale=0.5]{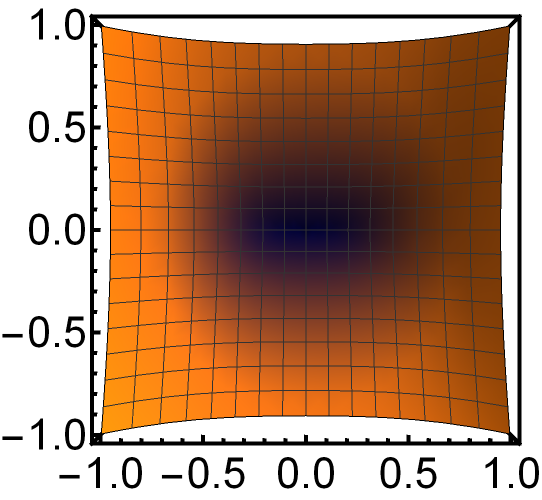}}
    \caption{(Color online) The cross-sectional plot of the energy $E = E(\zeta, k_{x}, k_{y})$ with the horizontal and
vertical axes respectively corresponding to $k_{x}$ and $k_{y}$ for fixing:
(a) $\zeta=0.1$, (b) $\zeta=0.5$, and (c) $\zeta=0.9$.}\label{fig.isenergetic.profile}
\end{figure*}

\section{Microscopic model and effective theory}\label{Sec_model}

The non-interacting Hamiltonian of a 2D SDSM with Dirac points takes the form of~\cite{Saha2016PRB,Montambaux}
\begin{align}
H^{}_{0} (\mathbf{k})= \alpha k_{x}^{2} \sigma^{}_{1}+v^{}_{F} k^{}_{y}\sigma^{}_{2},
\end{align}
where $\alpha$ is a microstructural parameter that measures the inverse of quasiparticle
mass along $x$ and $v_{F}$ represents the Fermi velocity along the $y$ direction, respectively.
Hereby, $\sigma^{}_0$ denotes the $2 \times 2$ identity matrix, and $\sigma^{}_i$ with
$i = 1,2,3$ characterize the Pauli matrices, which satisfy the algebra
$\{\sigma^{}_i, \sigma^{}_j\} = 2 \delta^{}_{ij}$.

Subsequently, we bring out a tilting term along $k_x$ direction described by $\zeta \alpha k_{x}^{2}$ to $H_0$
and are left with the free Hamiltonian of 2D tilted SDSM as follows
\begin{align}
H^{}_{0} (\mathbf{k})=\zeta \alpha k_{x}^{2}+\alpha k_{x}^{2}\sigma^{}_{1}+v^{}_{F} k^{}_{y}\sigma^{}_{2},\label{H0_2}
\end{align}
where $\zeta$ is a dimensionless tilting parameter. The energy eigenvalues of the system can be derived from Eq.~\eqref{H0_2}
\begin{align}
E^{}_{\pm}(\mathbf{k})=\zeta\alpha k_{x}^{2}\pm\sqrt{(\alpha k_{x}^{2})^2+(v^{}_{F} k^{}_{y})^2} . \label{Energy}
\end{align}
This indicates that the basic structure of Fermi surface is heavily dependent upon the value of $\zeta$.
Specifically, the 2D tilted SDSMs can be classified into two types~\cite{Lee2018PRB}:
Type-\uppercase\expandafter{\romannumeral 1} case owns a point-like Dirac point as long as $|\zeta|<1$ and instead Type-\uppercase\expandafter{\romannumeral 2} situation is equipped with certain open Fermi surface (such as straight lines)
at $|\zeta|>1$~\cite{Lee2018PRB,Lee2019PRB,Wang2023arXiv}.

Without loss generality, we from now on only focus on the type-I scenario.
With all these above in hand, the free effective action of a 2D tilted SDSM can be expressed as~\cite{Lee2018PRB, Uchoa1704.08780}
\begin{align}
S^{}_{0}=& \int^{+\infty}_{-\infty}\frac{\mathrm{d}\omega}{2\pi}
\int\frac{\mathrm{d}^2\mathbf{k}}{(2\pi)^2}
\psi^\dagger_{\sigma}(i\omega,\mathbf{k})(-i\omega+\zeta\alpha k_{x}^{2}\nonumber\\
&+\alpha k_{x}^{2}\sigma^{}_{1}
+v^{}_{F} k_{y}\sigma^{}_{2})\psi^{}_{\sigma}(i\omega,\mathbf{k}), \label{eq_S0}
\end{align}
where the spinor $\psi(i\omega,\mathbf{k})$ serves as the low-energy excitations from the point-like Dirac points. This
gives rise to the free fermionic propagator~\cite{Wang2018}
\begin{align}
G_0(i\omega,\mathbf{k})=\frac{1}{-i\omega+\zeta\alpha k_{x}^{2}+\alpha k_{x}^{2}\sigma^{}_{1}+v^{}_{F} k_{y}\sigma^{}_{2}},
\end{align}
which will be used to derive the one-loop corrections of Feynman diagrams.

\begin{figure*}[htbp]
\centering
\subfigure[ ]{\includegraphics[scale=0.22]{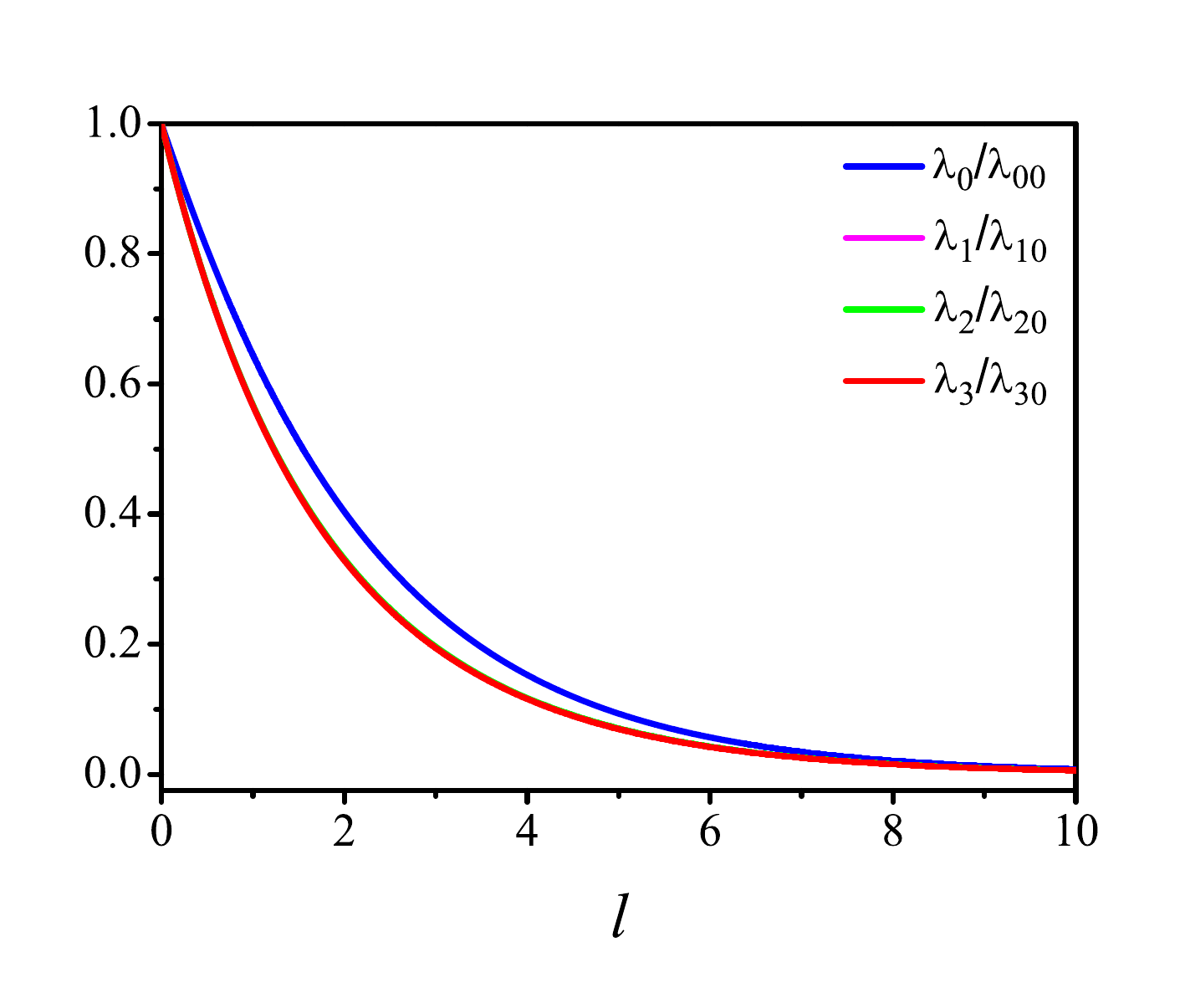}}
\hspace{-0.3cm}
\subfigure[ ]{\includegraphics[scale=0.22]{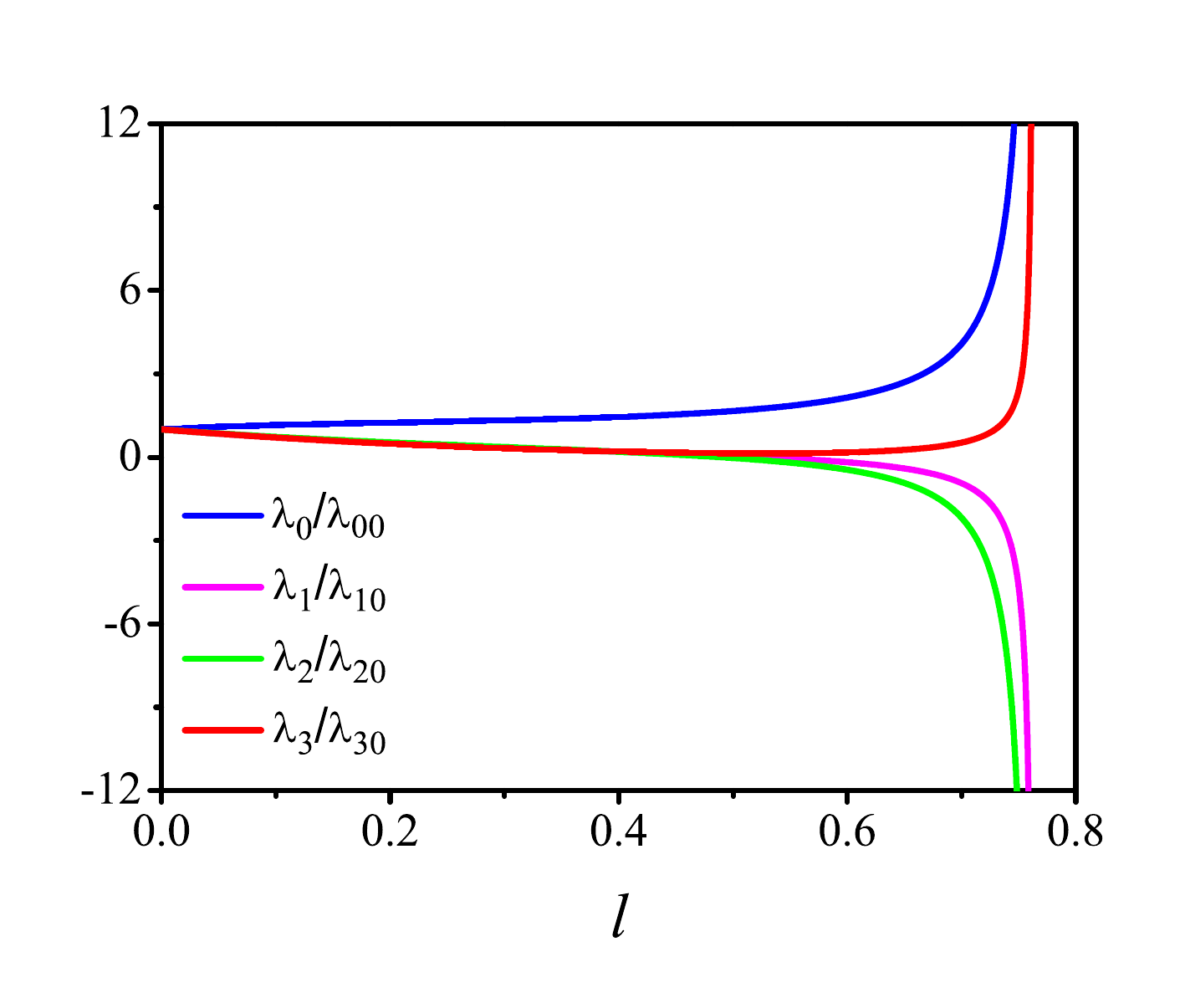}}
\hspace{-0.3cm}
\subfigure[ ]{\includegraphics[scale=0.22]{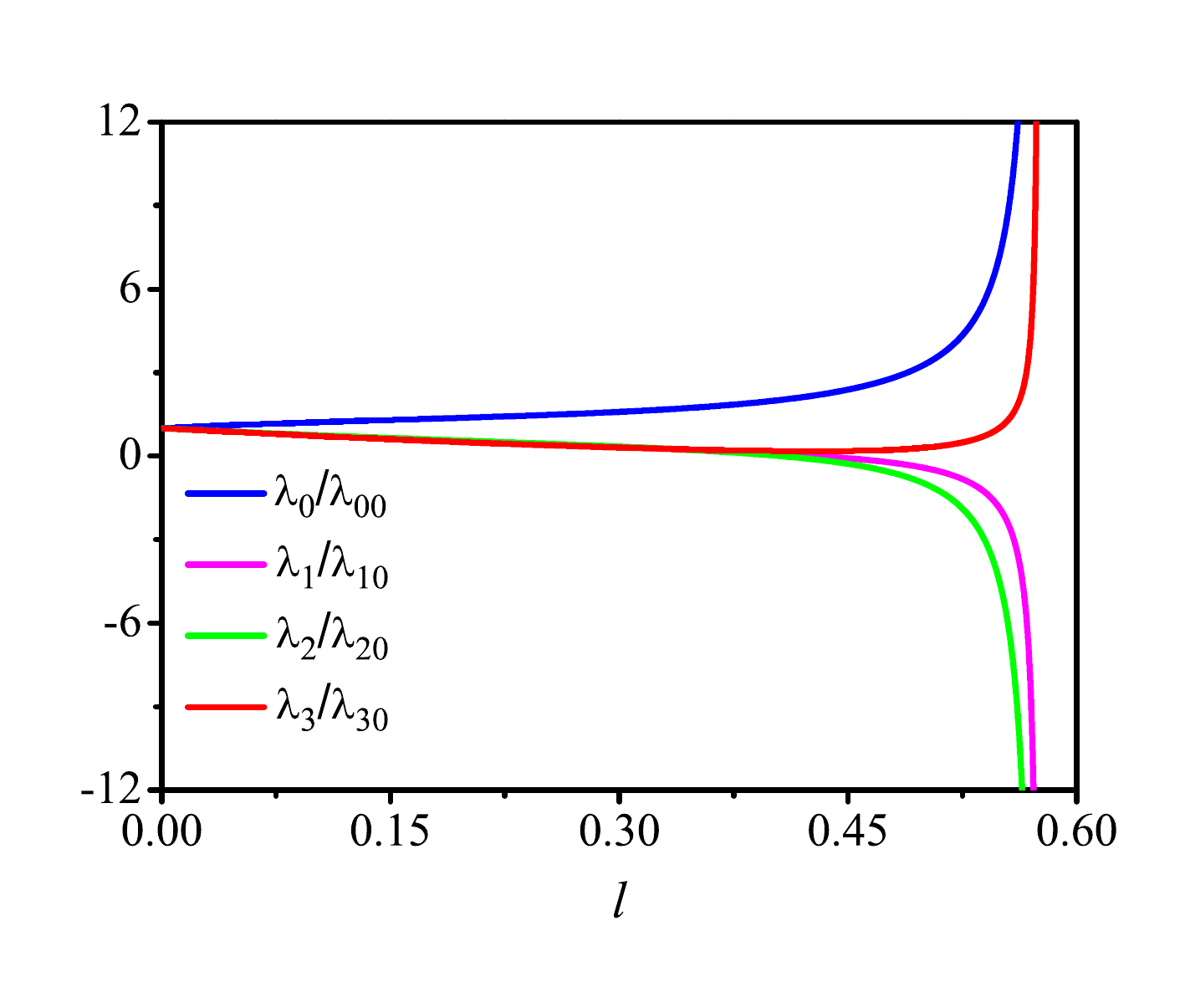}}
\caption{(Color online) Energy-dependent evolutions of fermion-fermion interaction
for both two Strategies: (a) $\zeta = 0.3$, $v_{F0} = \alpha_{0} =  10^{-3}$, and $\lambda_{i0} =  10^{-5}$ (Strategy I),
(b) $\zeta = 0.5$, $v_{F0} = 10^{-3}$, $\alpha_{0} =  10^{-4}$, and $\lambda_{i0} =  10^{-4}$ (Strategy  I),
(c) $\zeta=0.5$, $v_{F0} = 10^{-3}$ and $\alpha_{0} =  10^{-2}$ as well as $\lambda_{i0}=10^{-3}$ (Strategy  II).}\label{fig2}
\end{figure*}

Next, we introduce all possible fermion-fermion interactions that are allowed by the
free Hamiltonian~(\ref{eq_S0})~\cite{Lee2018PRB,Wang2018,Montambaux,Murray2014PRB}
\begin{widetext}
\begin{align}
S^{}_{\mathrm{int}}=\sum^3_{i=0}\lambda^{}_{i}\int\frac{\mathrm{d}\omega^{}_1 \mathrm{d}\omega^{}_2 \mathrm{d}\omega^{}_3}{(2\pi)^3}
\int\frac{
\mathrm{d}^2\mathbf{k}^{}_1
\mathrm{d}^2\mathbf{k}^{}_2
\mathrm{d}^2\mathbf{k}^{}_3}{(2\pi)^6} \Psi^\dag(\omega^{}_1,\mathbf{k}^{}_1)\sigma_i
\Psi(\omega^{}_2,\mathbf{k}^{}_2)\Psi^\dag(\omega^{}_3,\mathbf{k}^{}_3)\sigma^{}_i
\Psi(\omega^{}_1+\omega^{}_2-\omega^{}_3,\mathbf{k}^{}_1+\mathbf{k}^{}_2-\mathbf{k}^{}_3), \label{S_int}
\end{align}
\end{widetext}
where the matrix $\sigma_i$ and parameter $\lambda^{}_{i}$ with $i = 0, 1, 2, 3$ are adopted to
distinguish the different kinds of interactions and designate the strength of
related fermion-fermion interactions.
Eventually, we arrive at the low-energy effective theory after combining
the free term~(\ref{eq_S0}) and the interactions~(\ref{S_int})
\begin{align}
S_{\mathrm{eff}} =  S_{0}+S_{\mathrm{int}}.\label{Eq_S_eff}
\end{align}
Subsequently, we start from the effective action~(\ref{Eq_S_eff}) and perform all the one-loop corrections due to the fermion-fermion
interactions, with which our RG equations can be established. In this sense, we are able to
examine the critical behavior in the presence of fermion-fermion interactions in the low-energy regime. For simplicity,
we store all the detailed calculations for the one-loop corrections in Appendix~\ref{Appendix_1L-corrections}.

\section{RG analysis and coupled evolutions}\label{Sec_RGEqs}

In accordance with the spirit of RG approach~\cite{Wilson1975RMP,Polchinski1992arXiv,Shankar1994RMP}, we perform the RG process
by eliminating and reshaping the thin momentum shells to access the Fermi surface.
This leads to the coupled RG evolutions for relevant parameters in a given physical system.
However, the conventional momentum-shell RG approach is not suitable for the 2D tilted SDSMs owing to
its unique dispersion of low-energy excitations with being parabolic in one direction and
linear in the other. Instead, the energy on-shell RG method must be employed, which requires
us to integrate out a thin energy shell one by one during the RG analysis~\cite{Lee2018PRB,Lee2019PRB,Wang2018}.

In order to work in the energy-shell framework, we hereby adopt the strategies advocated in Refs.~\cite{Lee2018PRB,Lee2019PRB,Wang2018,Montambaux}. To this end, it is helpful to make
transformations from $k_x$ and $k_y$ to $E$ and $\theta$,
where $E$ and $\theta$ represent the energy eigenvalue and the
relative phase between $k_x$ and $k_y$, respectively. To be specific, we designate
\begin{align}
k^2_{x}  \equiv \frac{E (\cos \theta - \zeta)}{\alpha (1 - \zeta ^2)}, \,\,\,\,\,\,\,\,\,
k^{}_{y}  \equiv \frac{E \sin \theta}{v^{}_{F} \sqrt{1 - \zeta ^2}}, \label{HJ2}
\end{align}
where $\theta$ is restricted to the range of $-\pi \le \theta \le \pi$. This clusters into two
situations depending upon the sign of $E$, namely
$- \mathcal{X} < \theta < \mathcal{X} \,\,(E > 0)$ and
$- \pi < \theta < - \mathcal{X} \text{ and } \mathcal{X}< \theta <  \pi ,
\,\,(E < 0)$, with $\mathcal{X}\equiv\arccos \zeta$.
On the basis of these together with
$\mathrm{d} k^{}_x\,  \mathrm{d} k^{}_y = |J| \, \mathrm{d}E \, \mathrm{d} \theta$,
we are left with the following identities
\begin{align}
\int \!\!\mathrm{d} k^{}_{x}  \mathrm{d} k^{}_{y}
=\!\left \{
\begin{array} {l l}
\!\!4 \!\int_{0}^{\mathcal{X}} \!\mathrm{d} \theta\int_{0}^{\Lambda} \mathrm{d} E
\frac{1-\zeta \cos \theta}{2 v\left(1-\zeta^{2}\right)} \sqrt{\frac{E}{\alpha(\cos \theta-\zeta)}},  E > 0, \\ \\
\!\!4 \!\int_{\mathcal{X}}^{\pi} \mathrm{d} \theta \int_{-\Lambda}^{0} \mathrm{d} E
\frac{1-\zeta \cos \theta}{2 v\left(1-\zeta^{2}\right)} \sqrt{\frac{E}{\alpha(\cos \theta-\zeta)}},  E < 0,
\end{array} \right.\nonumber
\end{align}
with $\Lambda$ being a cutoff associated with the lattice constant. With these transformations in hand, we are able to evaluate the one-loop corrections shown in Figs.~\ref{fig10} and \ref{fig11}. After performing lengthy but straightforward calculations, following similar steps as in Refs.~\cite{Murray2014PRB,Wang2017PRB_QBCP,DZZW2020PRB,Sachdev2008PRB,Chubukov2010,Balatsky2010PRB,
JWangPRB2011,JWangPRB2013PRB,Kivelson2008PRB,Kim2015PRB,Sau2016PRB,Chubukov2016PRX} we can
obtain the one-loop contributions, which are provided in Appendix~\ref{Appendix_1L-corrections} for details.

Before proceeding further, we establish the non-interacting components of the effective action as a fixed point, ensuring their invariance throughout the RG transformations. This gives rise to the RG rescaling transformations of energies and momenta as well as fields in the momentum-frequency space~\cite{Lee2018PRB,Lee2019PRB,Wang2018},
\begin{eqnarray}
E&\rightarrow&E'=Ee^{-l},\label{Eq_scaling_1}\\
\omega&\rightarrow&\omega'=\omega e^{-l},\\
k_{x,y}&\rightarrow&k'_{x,y}=k_{x,y}e^{(-1-\eta )l},\\
\psi&\rightarrow&\psi'=\psi\ e^{2l+\eta l},\\
v&\rightarrow&v'=ve^{\eta l},\\
\alpha&\rightarrow&\alpha'=\alpha e^{(1+2\eta )l},\label{Eq_scaling_2}
\end{eqnarray}
where the variable parameter $l$ specifies an energy scale that is closely linked with
the cutoff, namely $E=\Lambda e^{-l}$ with $\Lambda$ being a cutoff.
Hereby, it is worth highlighting that we introduce a variable parameter $\eta\in(-1/2,0)$ to characterize the unique features of a 2D tilted SDSM, where $\eta=0$ and $\eta=-1/2$ correspond to the linear Dirac semimetals (SMs)~\cite{Neto2009RMP} and the quadratic-band-crossing point (QBCP) SM with a quadratic energy
dispersion~\cite{Kivelson2009PRL,Wang2017PRB_QBCP,Murray2014PRB,DZZW2020PRB,Fu2023arXiv}, respectively.
Further details concerning $\eta$ will be presented in Sec.~\ref{Sec_two_strategies}.
At current stage, we are now in a suitable position to carry out the RG analysis.
Following the standard procedures of RG approach~\cite{Shankar1994RMP,Lee2018PRB}, in conjunction with all the one-loop corrections provided in Appendix~\ref{Appendix_1L-corrections}, and utilizing the RG scaling transformations~(\ref{Eq_scaling_1})-(\ref{Eq_scaling_2}),
we arrive at the coupled flow RG equations of
all the interaction parameters in the effective theory~(\ref{Eq_S_eff}) as
\begin{widetext}

\vspace{-0.5cm}
\begin{align}
\frac{d v^{}_{F}}{d l} & =\eta v^{}_{F},  \label{RG_Eq.1}\\
\frac{d \alpha}{d l} & =(1 + 2 \eta) \alpha , \label{RG_Eq.2}\\
\frac{d \lambda^{}_{0}}{d l} & =\lambda^{}_{0}(-1-2 \eta) + \frac{1}{4 \pi^{2} v^{}_{F} \sqrt{\alpha}}
\frac{\left[\mathcal{F}^{}_{1}\left(\lambda_{0}^{2}+\lambda_{1}^{2}+\lambda_{2}^{2}+\lambda_{3}^{2}\right)-2 \mathcal{F}^{}_{2} \lambda^{}_{0} \lambda^{}_{1}-2 \mathcal{F}^{}_{3} \lambda^{}_{0} \lambda^{}_{2}\right]}{4 \pi^{2} v^{}_{F} \sqrt{\alpha}} , \label{RG_Eq.3} \\
\frac{d \lambda^{}_{1}}{d l} & =\lambda^{}_{1}(-1-2 \eta) + \frac{1}{4 \pi^{2} v^{}_{F} \sqrt{\alpha}}  \left[2 \mathcal{F}^{}_{1} \lambda^{}_{0} \lambda^{}_{1}-\mathcal{F}^{}_{2}\left(\lambda_{0}^{2}+\lambda_{1}^{2}+\lambda_{2}^{2}+\lambda_{3}^{2}\right)-2 \mathcal{F}^{}_{3}\left(2 \lambda_{1}^{2}+\lambda^{}_{0} \lambda^{}_{1}+\lambda^{}_{0} \lambda^{}_{3}-\lambda^{}_{1} \lambda^{}_{2}-\lambda^{}_{1} \lambda^{}_{3}\right)\right], \label{RG_Eq.4} \\
\frac{d \lambda^{}_{2}}{d l} & =\lambda^{}_{2}(-1-2 \eta)  + \frac{1}{4 \pi^{2} v^{}_{F} \sqrt{\alpha}}
\left[2 \mathcal{F}^{}_{1} \lambda^{}_{0} \lambda^{}_{2}-2 \mathcal{F}^{}_{2}\left(2 \lambda_{2}^{2}+\lambda^{}_{0} \lambda^{}_{2}+\lambda^{}_{0} \lambda^{}_{3}-\lambda^{}_{1} \lambda^{}_{2}-\lambda^{}_{2} \lambda^{}_{3}\right)-\mathcal{F}^{}_{3}\left(\lambda_{0}^{2}+\lambda_{1}^{2}+\lambda_{2}^{2}+\lambda_{3}^{2}\right)\right] , \label{RG_Eq.5} \\
\frac{d \lambda^{}_{3}}{d l} & =\lambda^{}_{3}(-1-2 \eta)+ \frac{1}{4 \pi^{2} v^{}_{F} \sqrt{\alpha}}
\left[2 \mathcal{F}^{}_{1}\left(-2 \lambda_{3}^{2}+\lambda^{}_{1} \lambda^{}_{3}+\lambda^{}_{2} \lambda^{}_{3}\right)-2 \mathcal{F}^{}_{2} \lambda^{}_{0} \lambda^{}_{2}-2 \mathcal{F}^{}_{3} \lambda^{}_{0} \lambda^{}_{1}\right], \label{RG_Eq.6}
\end{align}
\end{widetext}
where the coefficients are denominated as
\begin{eqnarray}
\mathcal{F}^{}_{1} & \equiv& \int_{0}^{\mathcal{X}} \!\!\!\!\mathrm{d} \theta \sqrt{\frac{1}{\cos \theta-\zeta}}
+\int_{\mathcal{X}}^{\pi} \!\!\!\!\mathrm{d}  \theta\,  \sqrt{\frac{1}{\zeta-\cos \theta}},  \label{F_1}\\
\mathcal{F}^{}_{2} & \equiv& \int_{0}^{\mathcal{X}} \!\!\!\!\mathrm{d} \theta \frac{\sqrt{(\cos \theta-\zeta)^{3}}}{(1-\zeta \cos \theta)^{2}}
+\int_{\mathcal{X}}^{\pi} \!\!\!\!\mathrm{d} \theta   \frac{\sqrt{(\zeta-\cos \theta)^{3}}}{(1-\zeta \cos \theta)^{2}} ,  \label{F_2}\\
\mathcal{F}^{}_{3} & \equiv& \int_{0}^{\mathcal{X}} \!\!\!\!\mathrm{d} \theta \frac{\sin ^{2} \theta\left(1-\zeta^{2}\right)}{(1-\zeta \cos \theta)^{2} \sqrt{\cos \theta-\zeta}}\nonumber\\
&&+\int_{\mathcal{X}}^{\pi} \!\!\!\!\mathrm{d} \theta  \frac{\sin ^{2} \theta\left(1-\zeta^{2}\right)}{(1-\zeta \cos \theta)^{2} \sqrt{\zeta-\cos \theta}},  \label{F_3}
\end{eqnarray}
with $\mathcal{X}\equiv\arccos \zeta$. These equations reveal that the four-fermion interaction parameters ($\lambda^{}_{0}, \lambda^{}_{1},\lambda^{}_{2},\lambda^{}_{3}$), and the fermion velocity ($v^{}_{F}$), as well as the
microstructure parameter ($\alpha$) are entangled with each other.
Consequently, one can expect unusual behavior in the low-energy regime, which will be discussed
in detail in Sec.~\ref{Sec_phyiscal_behavior}.

\begin{figure}[htbp]
\centering
\subfigure[ ]{\includegraphics[scale=0.22]{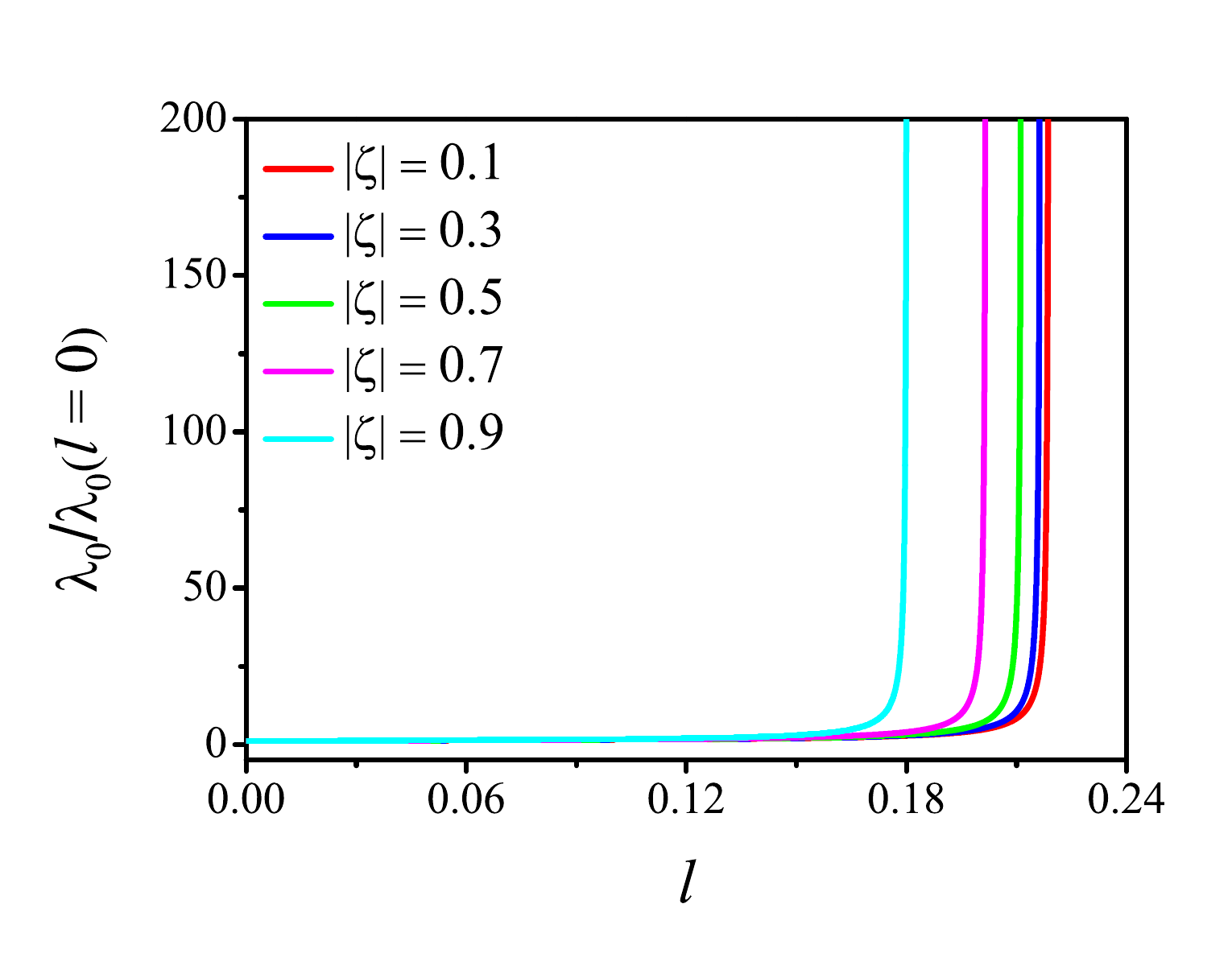}}
\vskip -0.2cm
\subfigure[ ]{\includegraphics[scale=0.22]{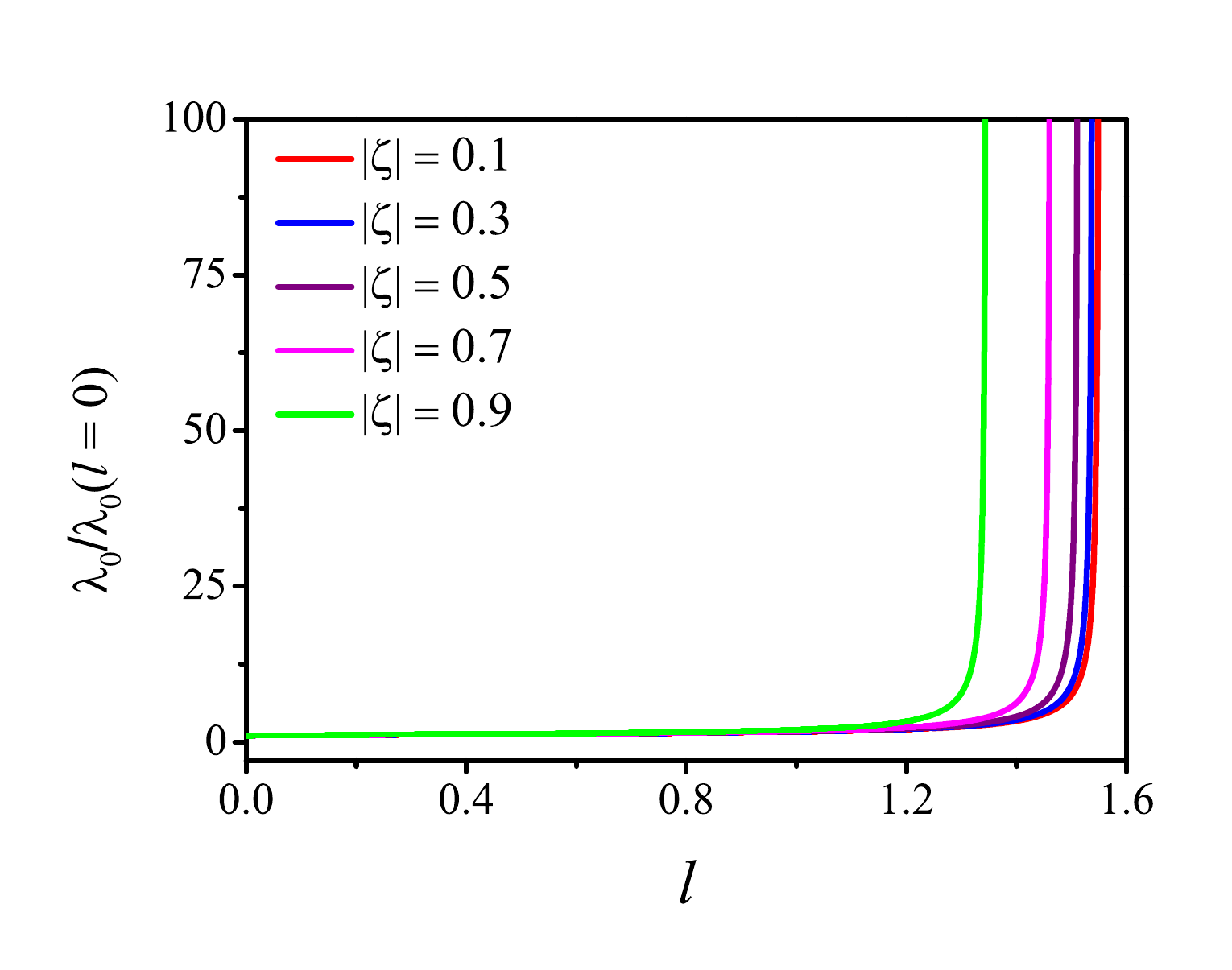}}
\caption{(Color online) Energy-dependent evolutions of $\lambda_0/\lambda_0(l=0)$ for both two Strategies: (a) $v_{F0}=10^{-4}$, $\alpha_{0}=10^{-3}$, and $\lambda_{i0}=10^{-4}$ (Strategy I), and (b) $v_{F0}=10^{-4}$, $\alpha_{0}=10^{-3}$, and $\lambda_{i0}=10^{-5}$ (Strategy II).}\label{fig3}
\end{figure}

\begin{figure*}[htbp]
\centering
\subfigure[ ]{\includegraphics[scale=0.21]{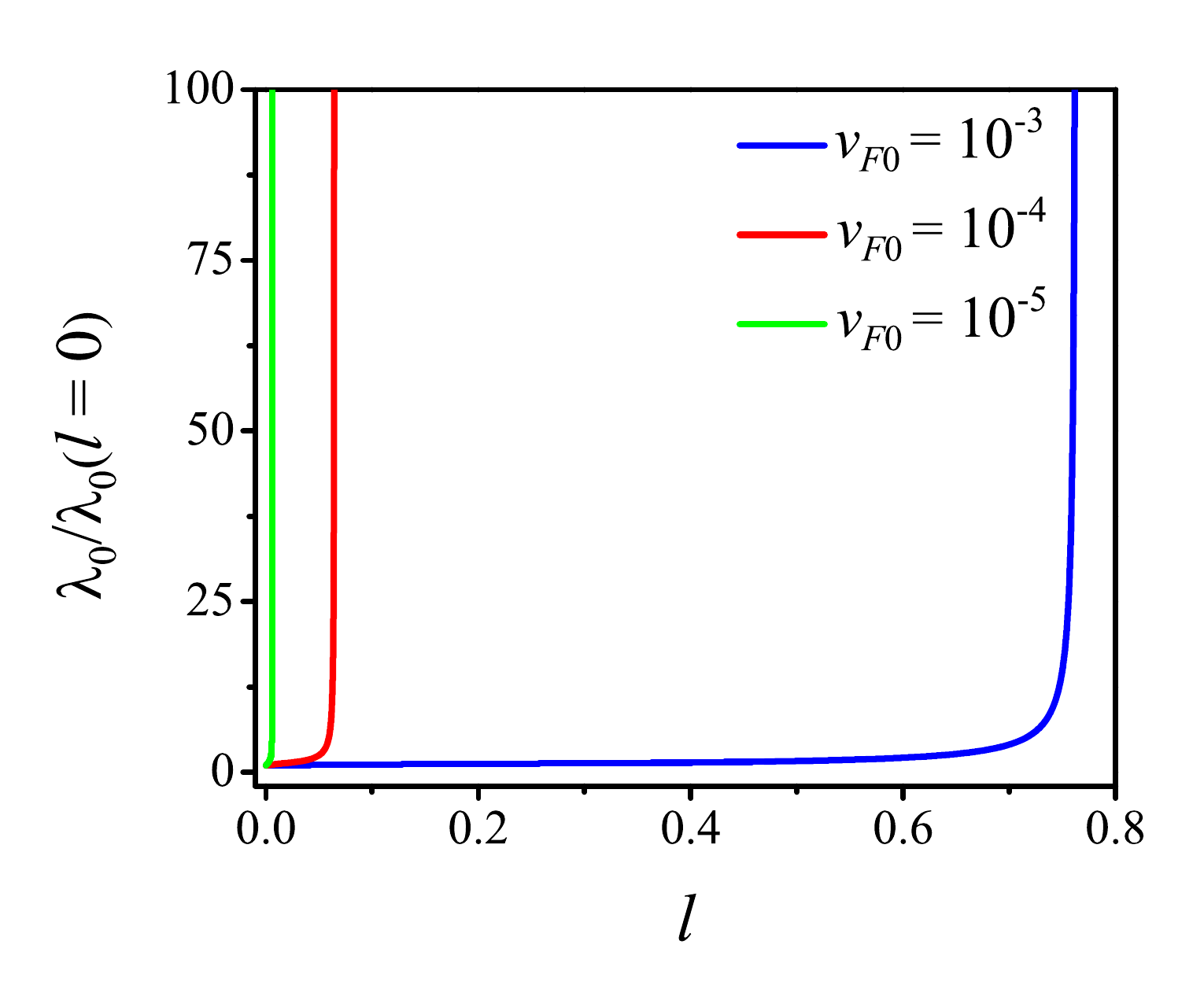}}
\hspace{-0.3cm}
\subfigure[ ]{\includegraphics[scale=0.21]{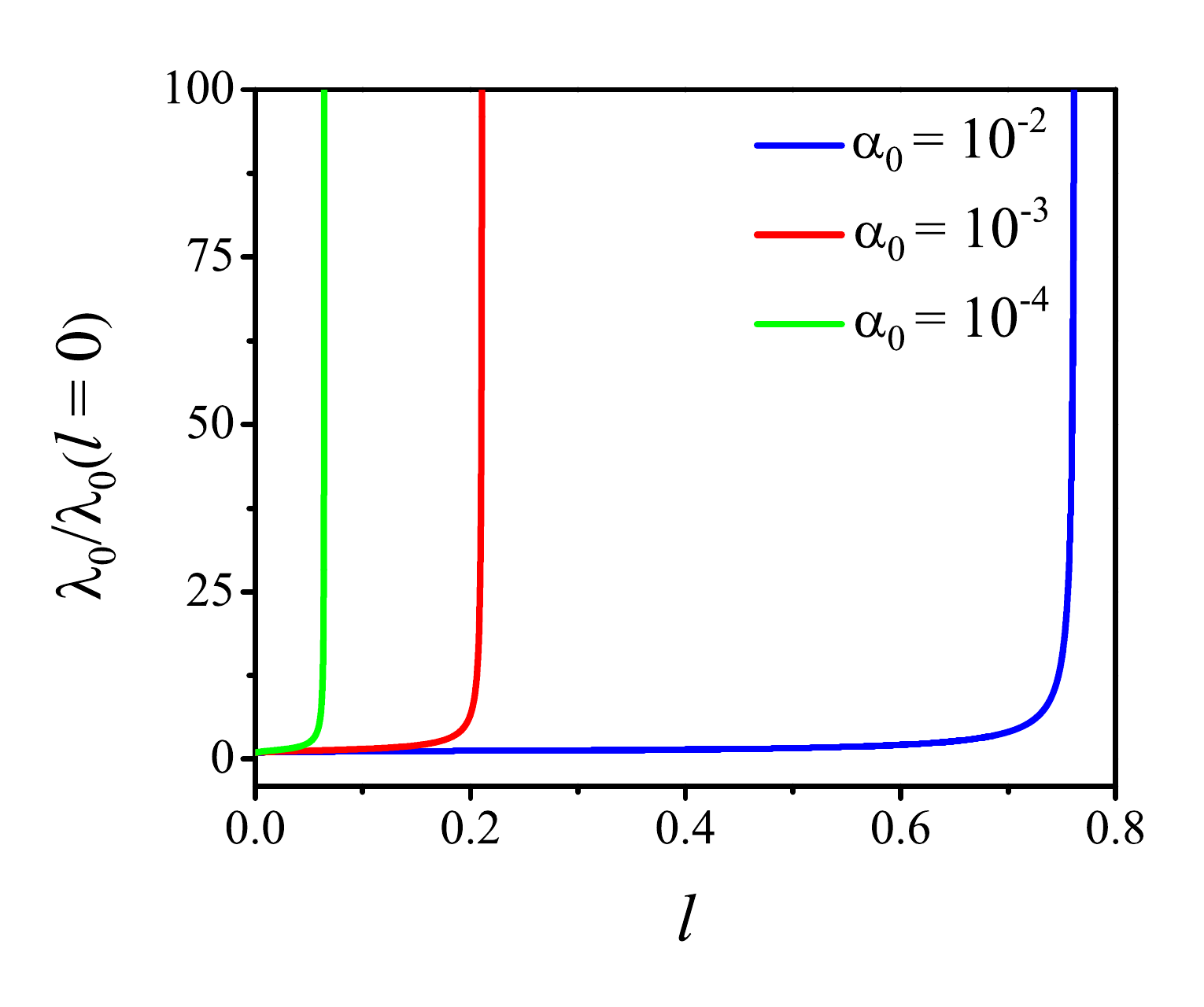}}
\hspace{-0.3cm}
\subfigure[ ]{\includegraphics[scale=0.21]{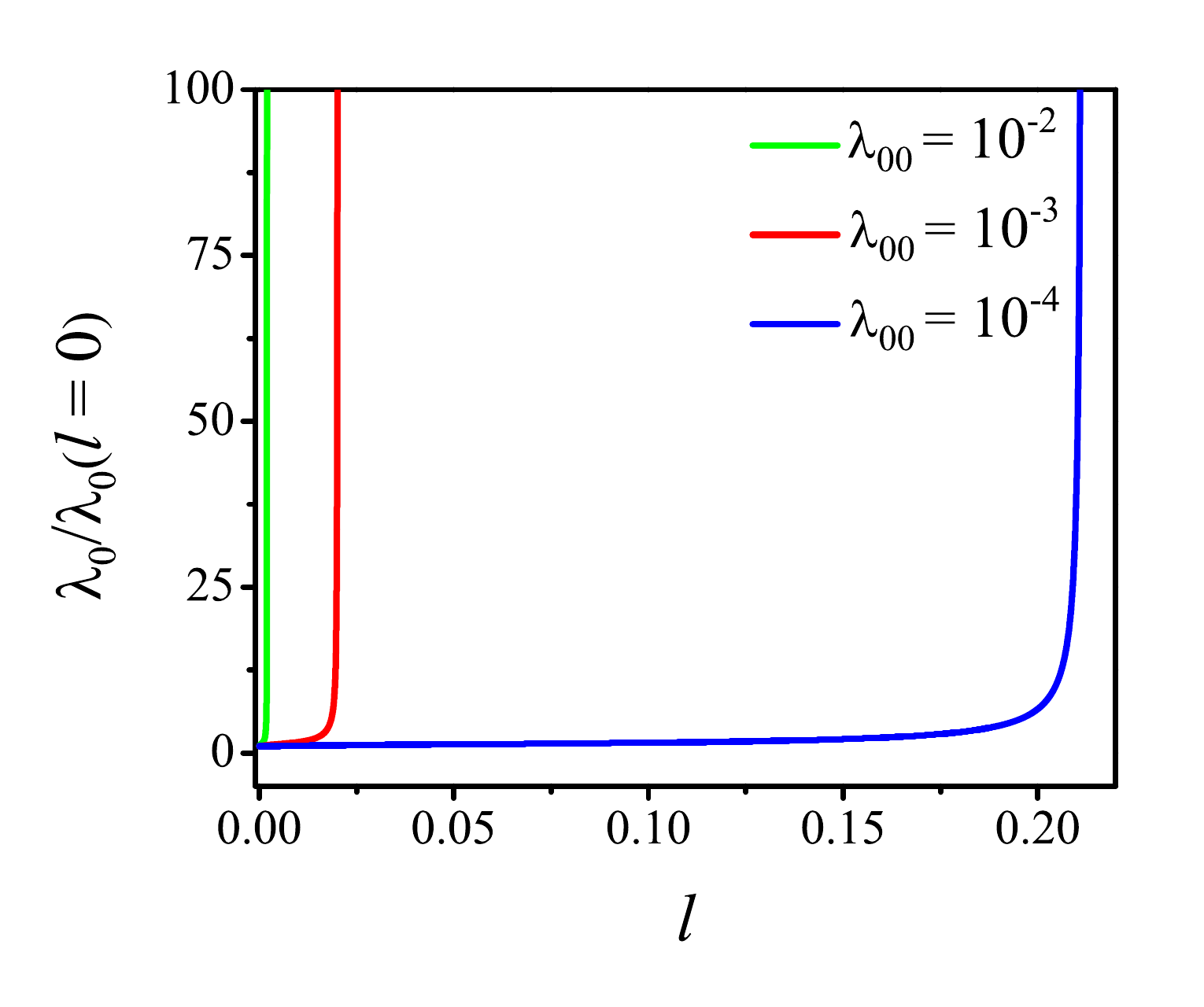}}\\\vspace{-0.1cm}
\subfigure[ ]{\includegraphics[scale=0.21]{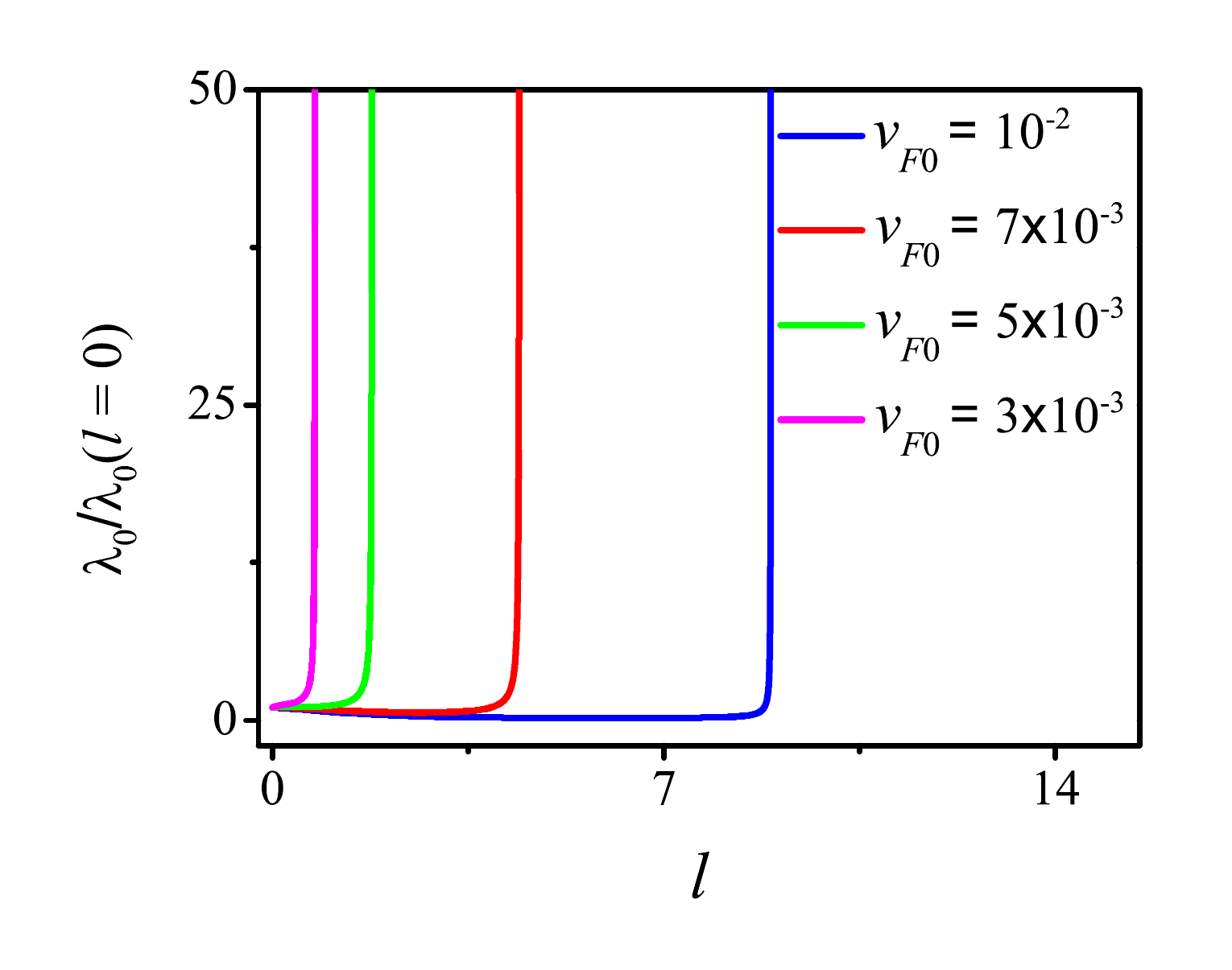}}
\hspace{-0.3cm}
\subfigure[ ]{\includegraphics[scale=0.21]{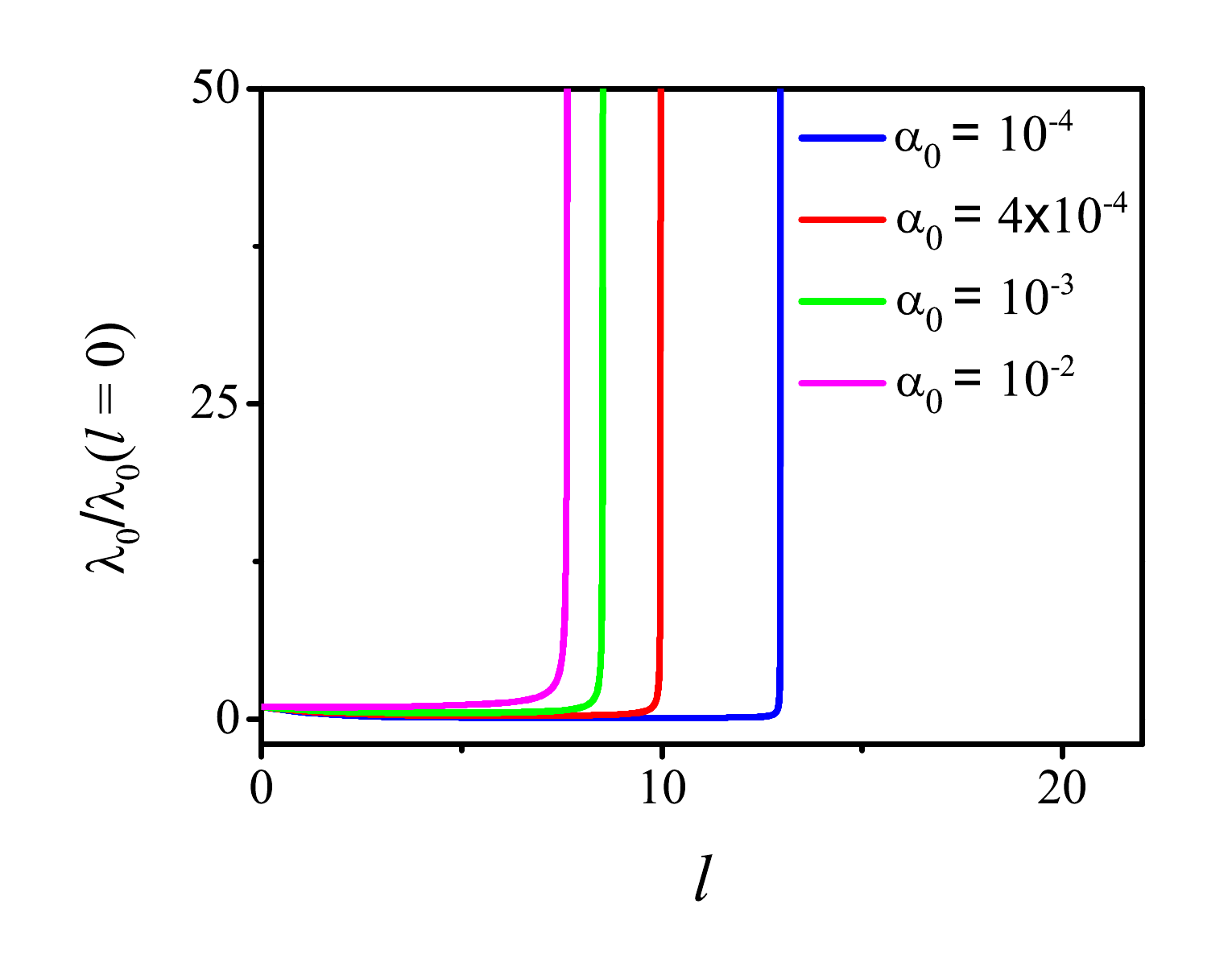}}
\hspace{-0.3cm}
\subfigure[ ]{\includegraphics[scale=0.21]{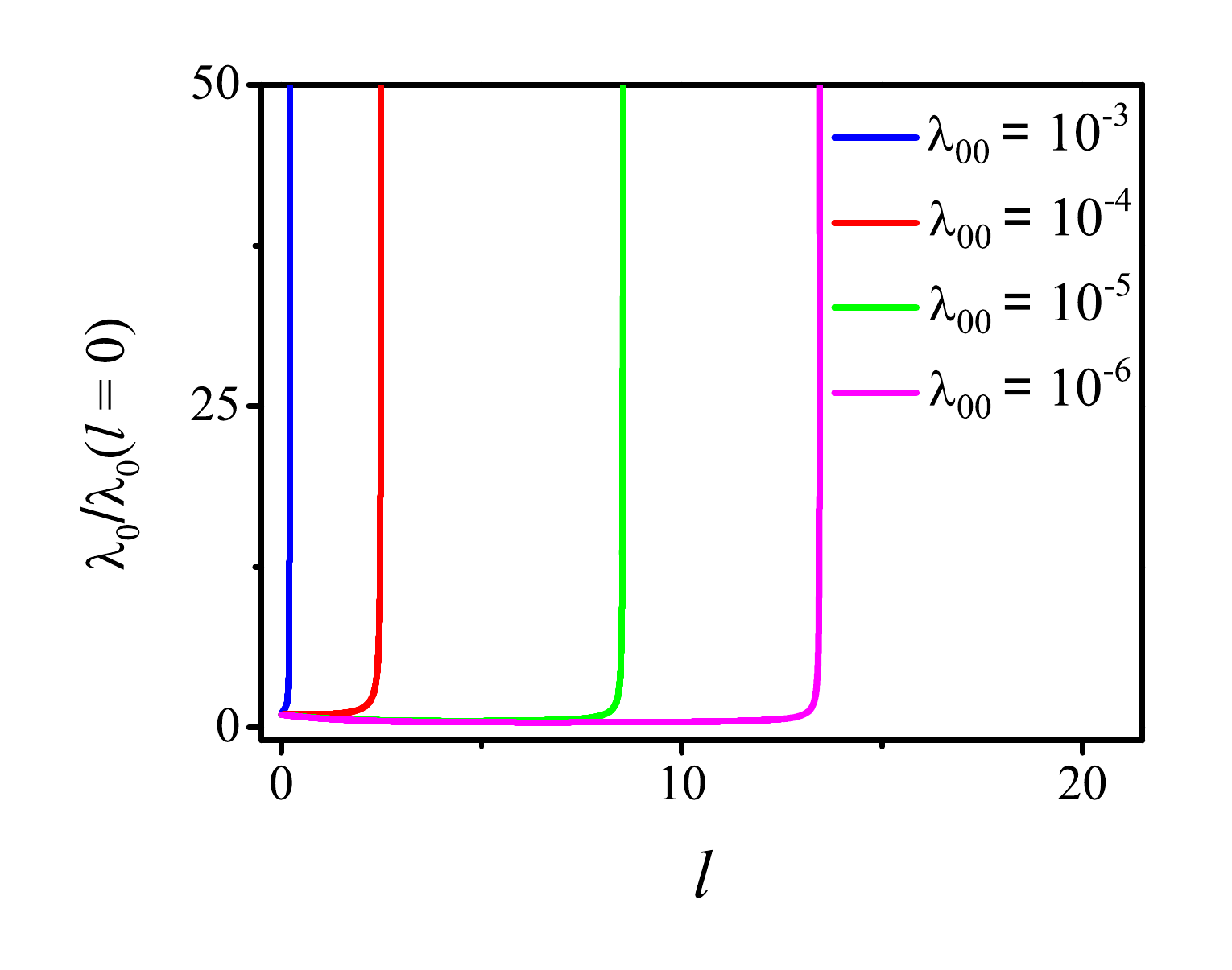}}
\caption{(Color online) Energy-dependent evolutions of $\lambda_0/\lambda_0(l=0)$ for (a) $v_{F0}$ with $\alpha_{0} = \lambda_{i0} = 10^{-4}$, (b) $\alpha_0$ with $v_{F0} = \lambda_{i0} = 10^{-3}$ and (c) $\lambda_{00}$ with $v_{F0} = 10^{-4}$, and $\alpha_{0} = 10^{-3}$ for (Strategy I), and (d) $v_{F0}$ with $\alpha_{0} = \lambda_{i0} = 10^{-3}$, (e) $\alpha_0$ with $v_{F0} = 10^{-3}$ and $\lambda_{i0} = 10^{-5}$ and (f) $\lambda_{00}$ with $v_{F0} = \alpha_{0} = 10^{-3}$ for (Strategy II) at $\zeta=0.5$.}
\label{fig4}
\end{figure*}

\section{Two distinct strategies for the parameter $\eta$}\label{Sec_two_strategies}

As aforementioned in Sec.~\ref{Sec_RGEqs}, a scaling parameter $\eta$ has been
introduced to construct the RG transformations that carry the features of a 2D tilted SDSM.
Although the exact value of $\eta$ is unclear, we stop here and turn our attention to the potential
expressions of this very parameter. Due to the unique dispersion of the 2D tilted SDSMs, $\eta$ is confined
within the range $\eta \in (-1/2, 0)$, with the boundaries $\eta=0$ and $\eta=-1/2$ corresponding to
the linear Dirac SMs~\cite{Neto2009RMP} and the QBCP SMs~\cite{Kivelson2009PRL,Wang2017PRB_QBCP,
Murray2014PRB,DZZW2020PRB,Fu2023arXiv}, respectively.
To simplify our analysis, it is helpful to consider
two distinct situations including Strategy \uppercase\expandafter{\romannumeral 1} with a constant $\eta$ and
Strategy \uppercase\expandafter{\romannumeral 2} where $\eta$ is also
dependent upon other parameters, i.e., a parameter-dependent $\eta$.

At first, we consider the Strategy \uppercase\expandafter{\romannumeral 1}.
The dispersion of a 2D tilted SDSM is an intermediate between that of Dirac and QBCP semimetal,
which corresponds to $\eta = 0$ and $\eta = -0.5$, respectively.
As a warm-up, we propose an ansatz that the contribution from the quadratic term of $k_{x}$ is
approximately equivalent to its linear-$k_{y}$ counterpart, giving rise to $\eta \approx -0.25$, which would roughly capture the main physics despite the exact value of $\eta$ remaining unknown at present. With the help of this concrete value
of parameter $\eta$, we can make a numerical analysis of RG equations~(\ref{RG_Eq.1})-(\ref{RG_Eq.6}), which will be clearly
addressed in Sec.~\ref{Sec_phyiscal_behavior}.

\begin{figure*}[htbp]
    \centering
    \subfigure[]{\includegraphics[scale=0.21]{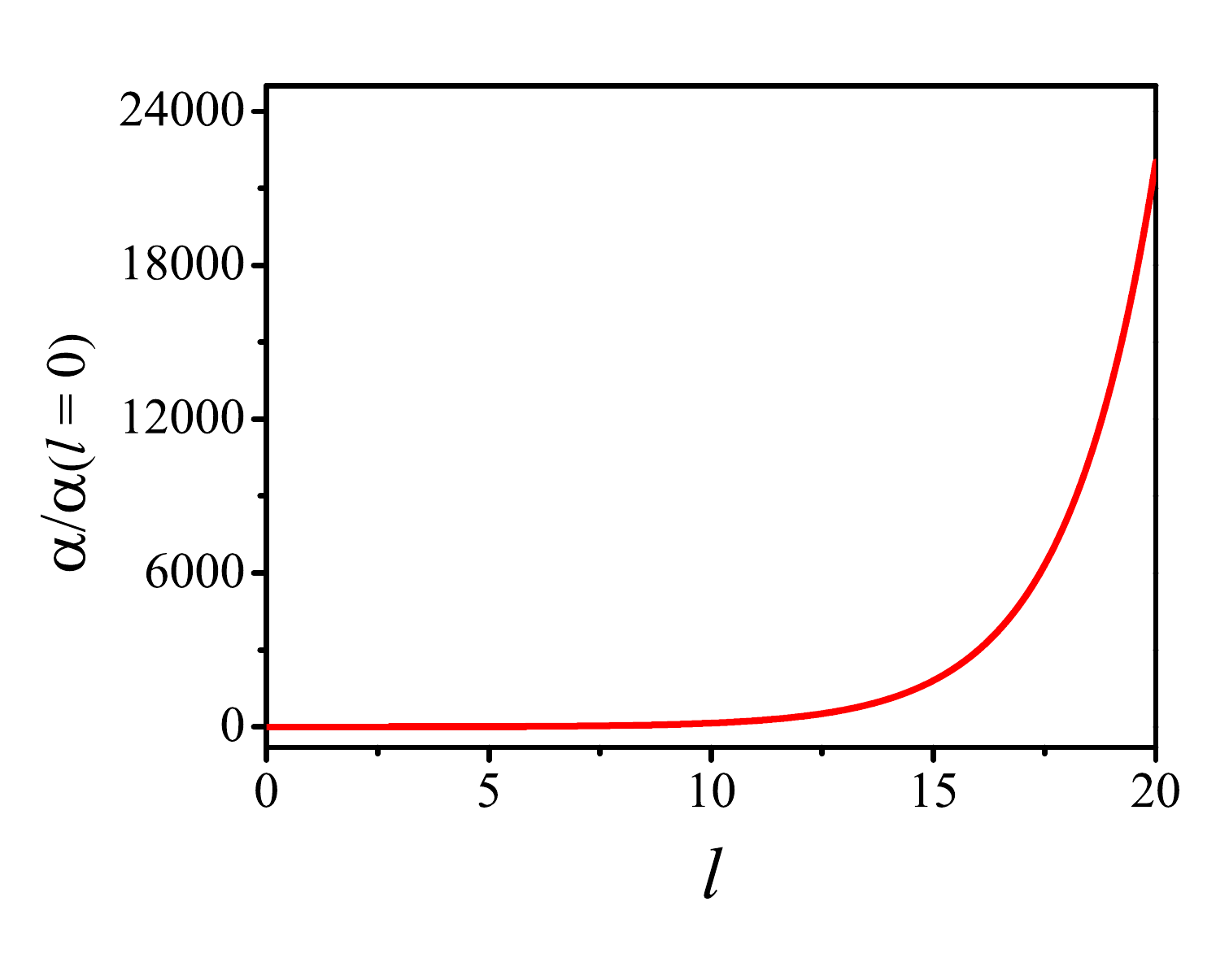}}
    \hspace{-0.3cm}
    \subfigure[]{\includegraphics[scale=0.21]{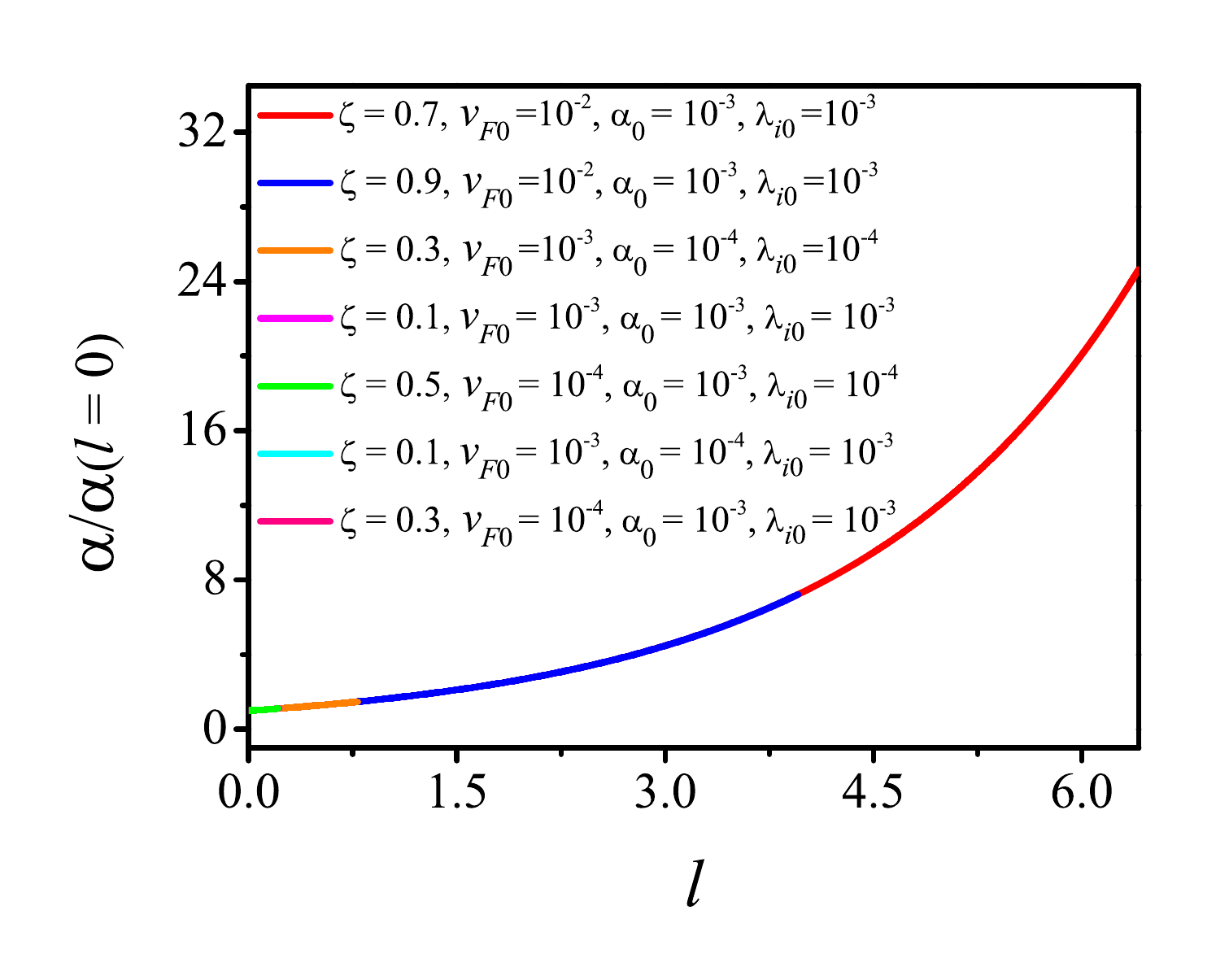}}
    \hspace{-0.3cm}
    \subfigure[]{\includegraphics[scale=0.21]{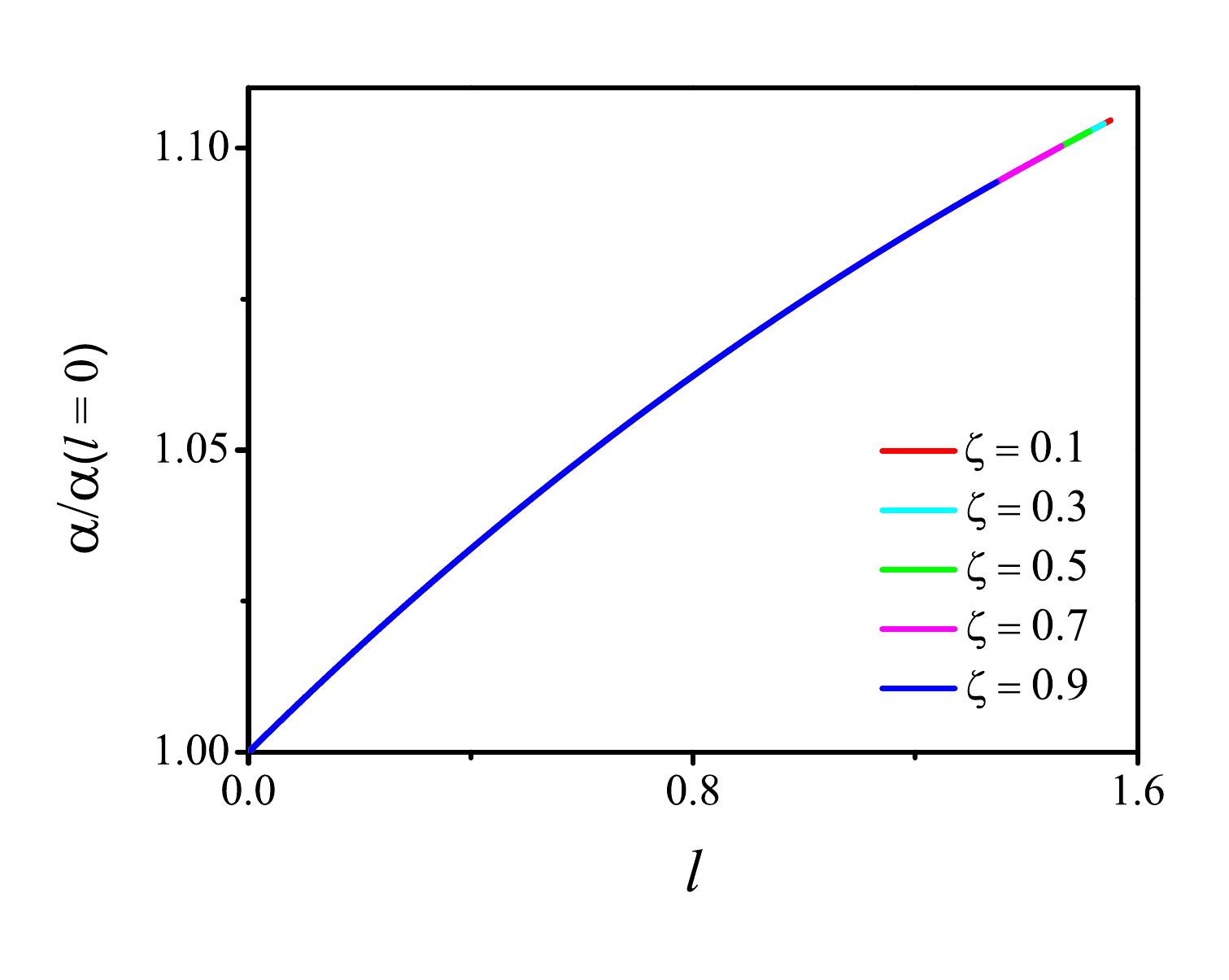}}\\ \vspace{-0.1cm}
    \subfigure[]{\includegraphics[scale=0.21]{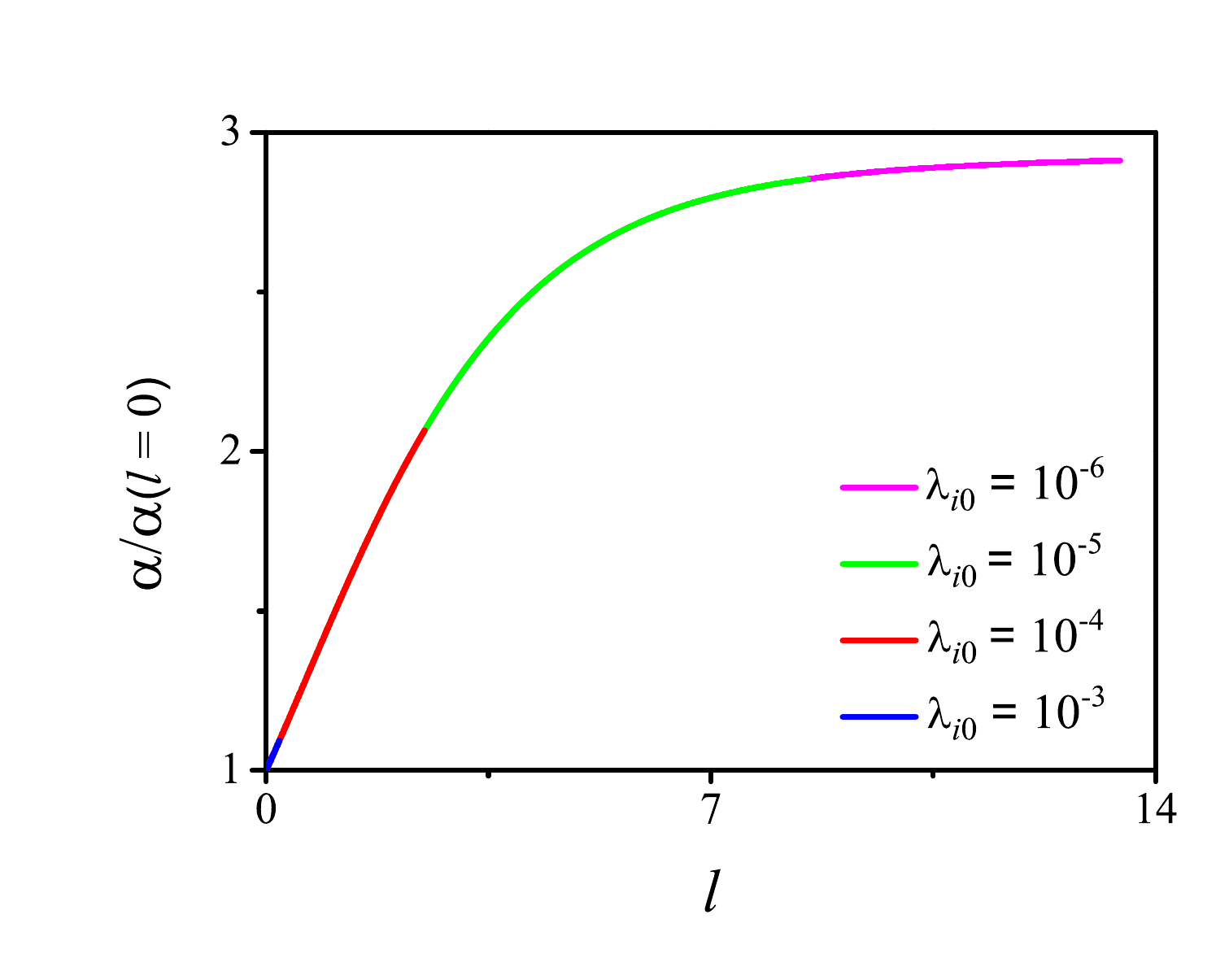}}
        \hspace{-0.3cm}
    \subfigure[]{\includegraphics[scale=0.21]{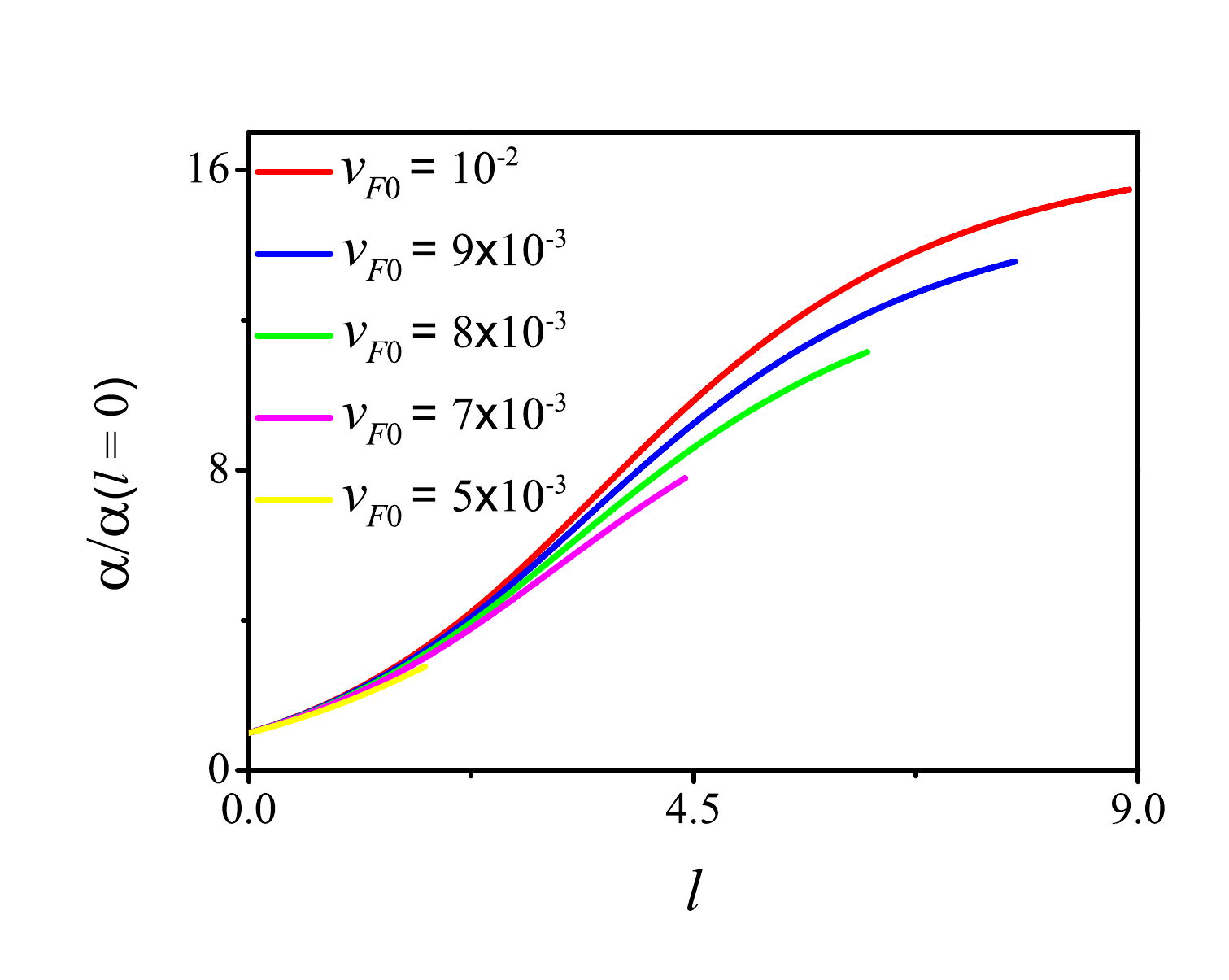}}
        \hspace{-0.3cm}
    \subfigure[]{\includegraphics[scale=0.21]{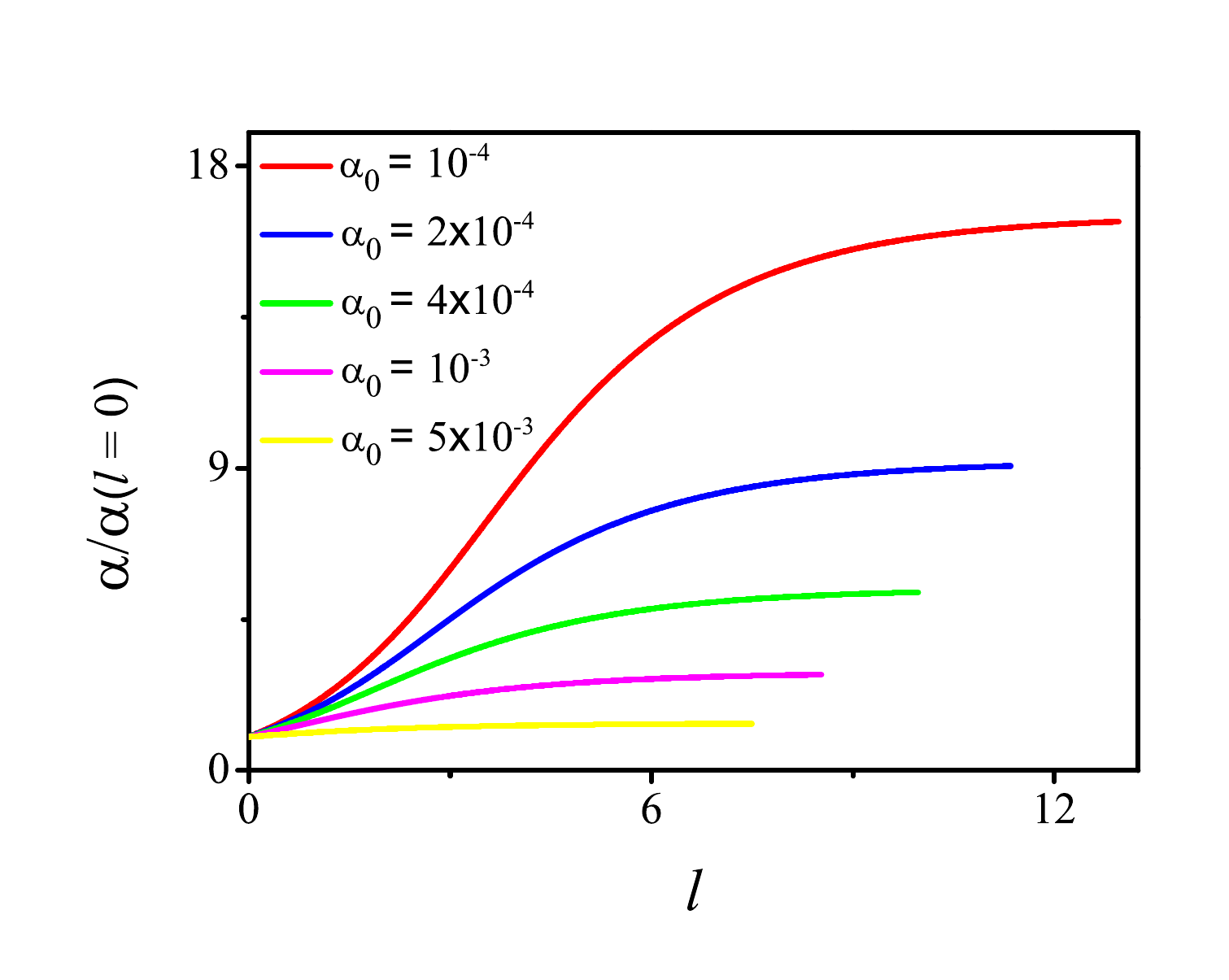}}
    \caption{(Color online) Energy-dependent evolutions of $\alpha/\alpha(l=0)$ for both two strategies:
(a) $\zeta = 0.5$, $v_{F0} = 10^{-3}$, $\alpha_{0} = 10^{-4}$,
and $\lambda_{i0} = 10^{-5}$ (Strategy \uppercase\expandafter{\romannumeral 1}),
(b) Various sets of $\zeta$, $v_{F0}$, $\alpha_{0}$, and $\lambda_{i0}$ (Strategy \uppercase\expandafter{\romannumeral 1}),
(c) $\zeta$ with $v_{F0} = 10^{-4}$, $\alpha_{0} = 10^{-3}$,
and $\lambda_{i0} = 10^{-5}$ (Strategy \uppercase\expandafter{\romannumeral 2}),
(d) $\lambda_{i0}$ with $\zeta = 0.5$, $v_{F0} = 10^{-3}$,
and $\alpha_{0} = 10^{-3}$ (Strategy \uppercase\expandafter{\romannumeral 2}),
(e) $v_{F0}$ with $\zeta = 0.5$, $\alpha_{0} = 10^{-3}$,
and $\lambda_{i0} = 10^{-3}$ (Strategy \uppercase\expandafter{\romannumeral 2}),
and (f) $\alpha_{0}$ with $\zeta = 0.5$, $v_{F0} = 10^{-3}$,
and $\lambda_{i0} = 10^{-5}$ (Strategy \uppercase\expandafter{\romannumeral 2}).}\label{fig5}
\end{figure*}

\begin{figure*}[htbp]
    \centering
    \subfigure[ ]{\includegraphics[scale=0.21]{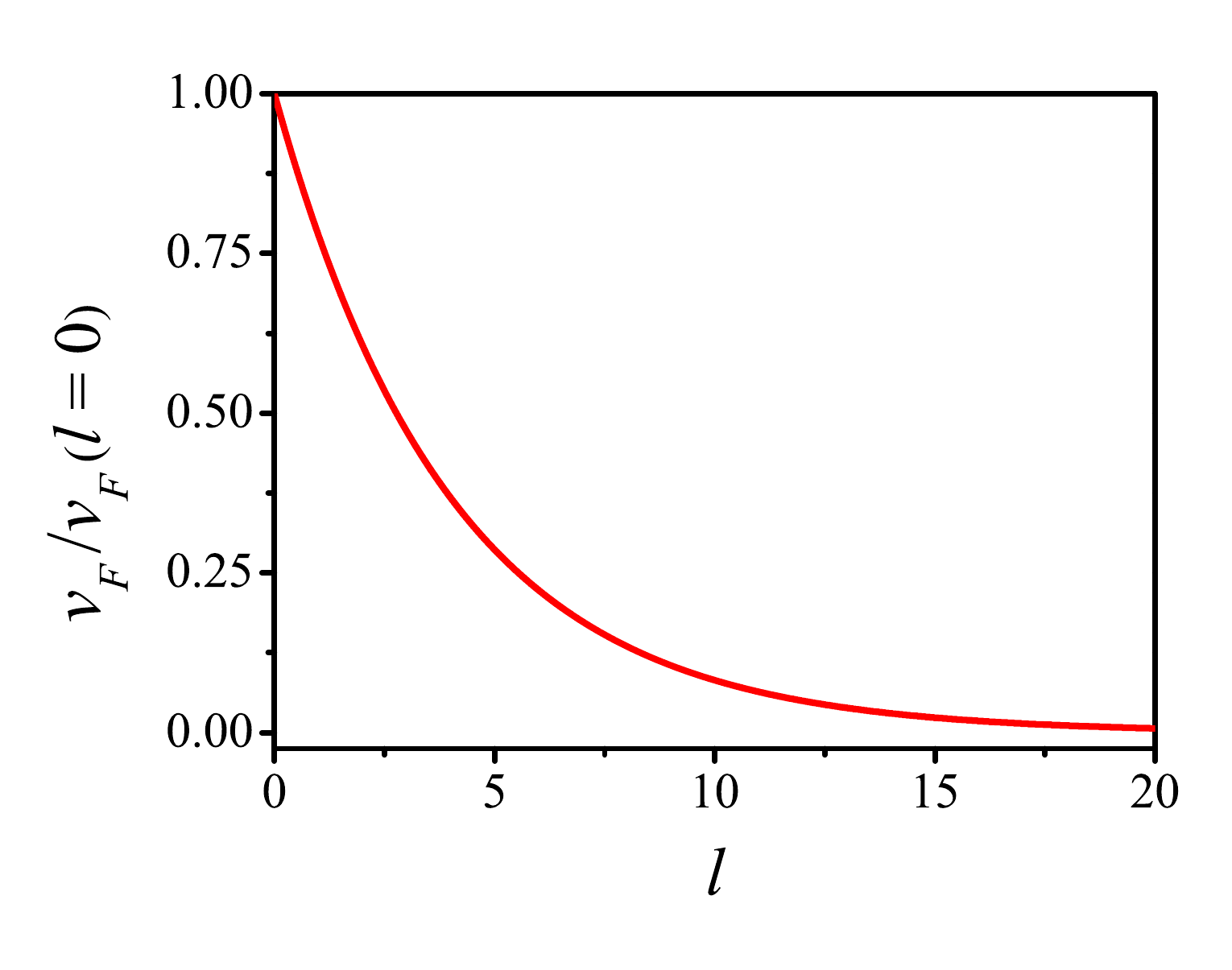}}
   \hspace{-0.3cm}
    \subfigure[ ]{\includegraphics[scale=0.21]{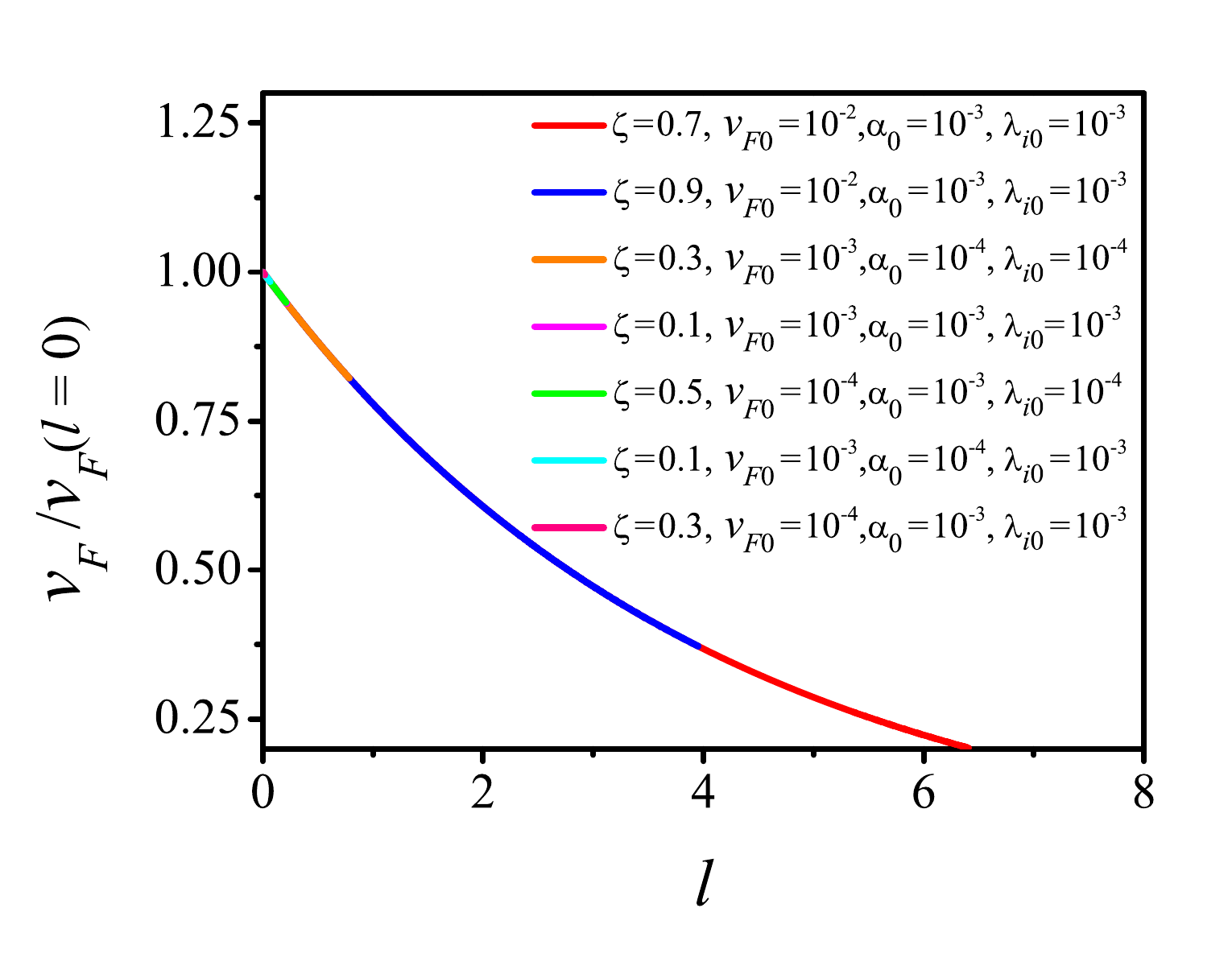}}
    \hspace{-0.3cm}
    \subfigure[ ]{\includegraphics[scale=0.21]{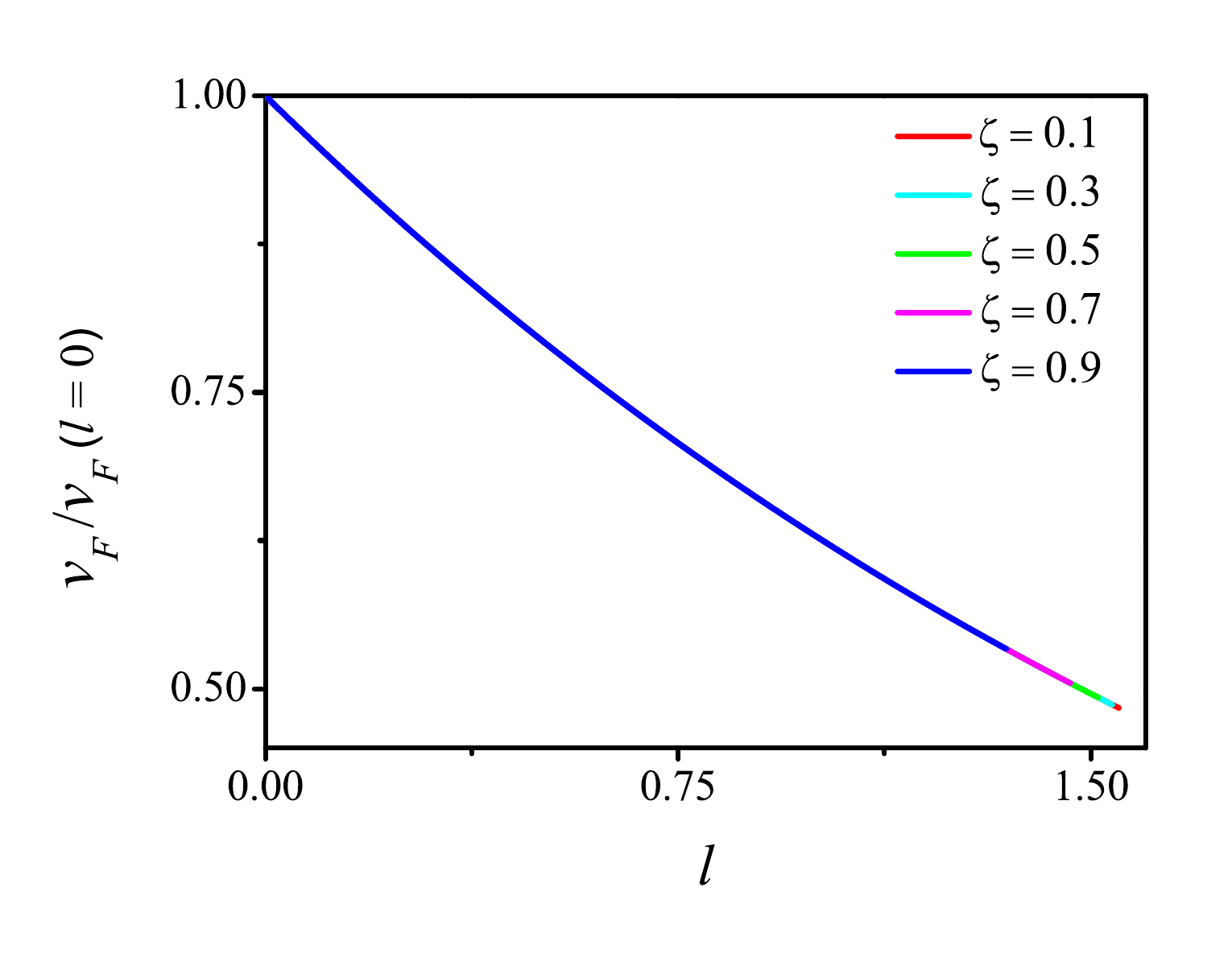}}\\\vspace{-0.1cm}
    \subfigure[ ]{\includegraphics[scale=0.21]{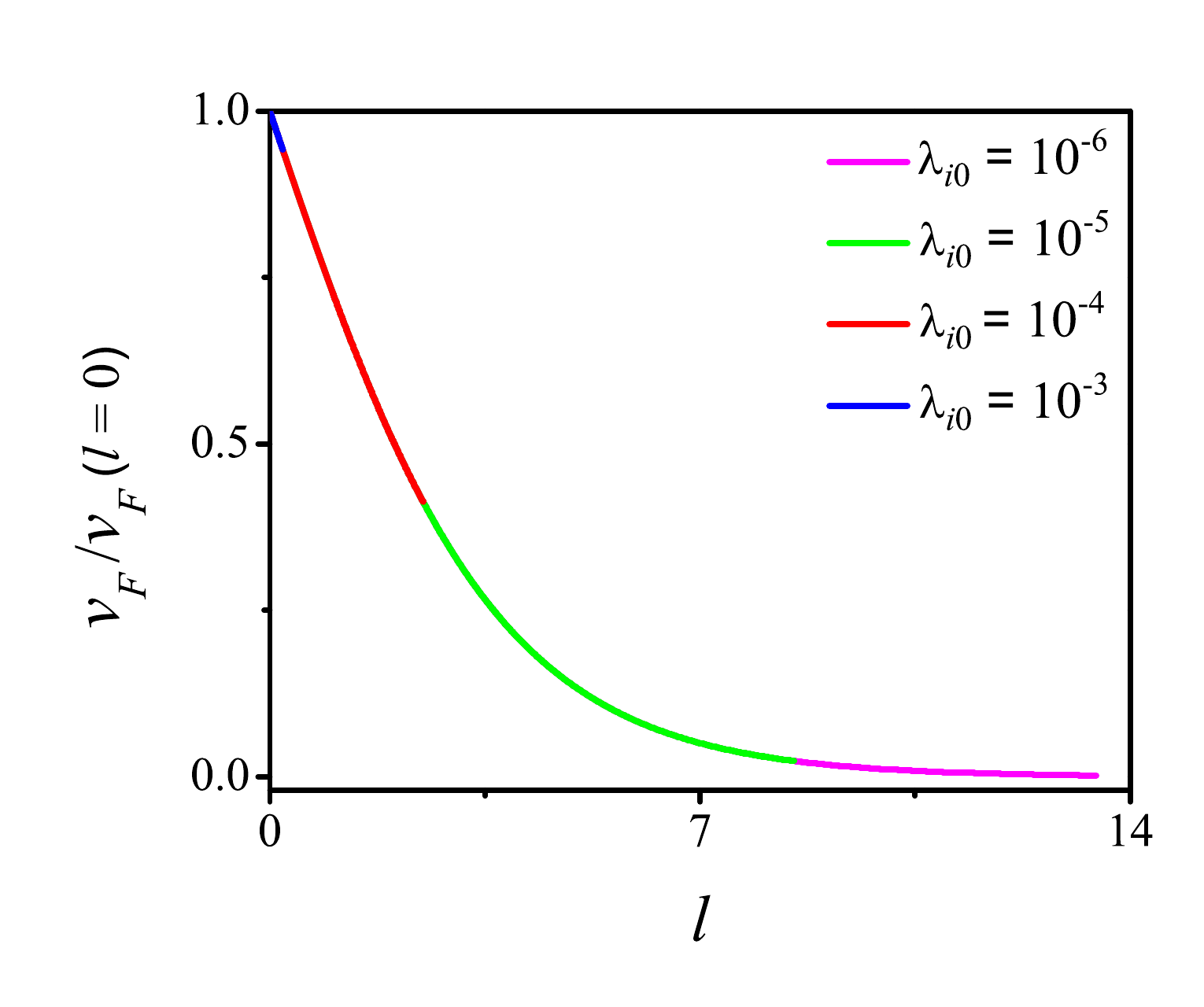}}
    \hspace{-0.3cm}
    \subfigure[ ]{\includegraphics[scale=0.21]{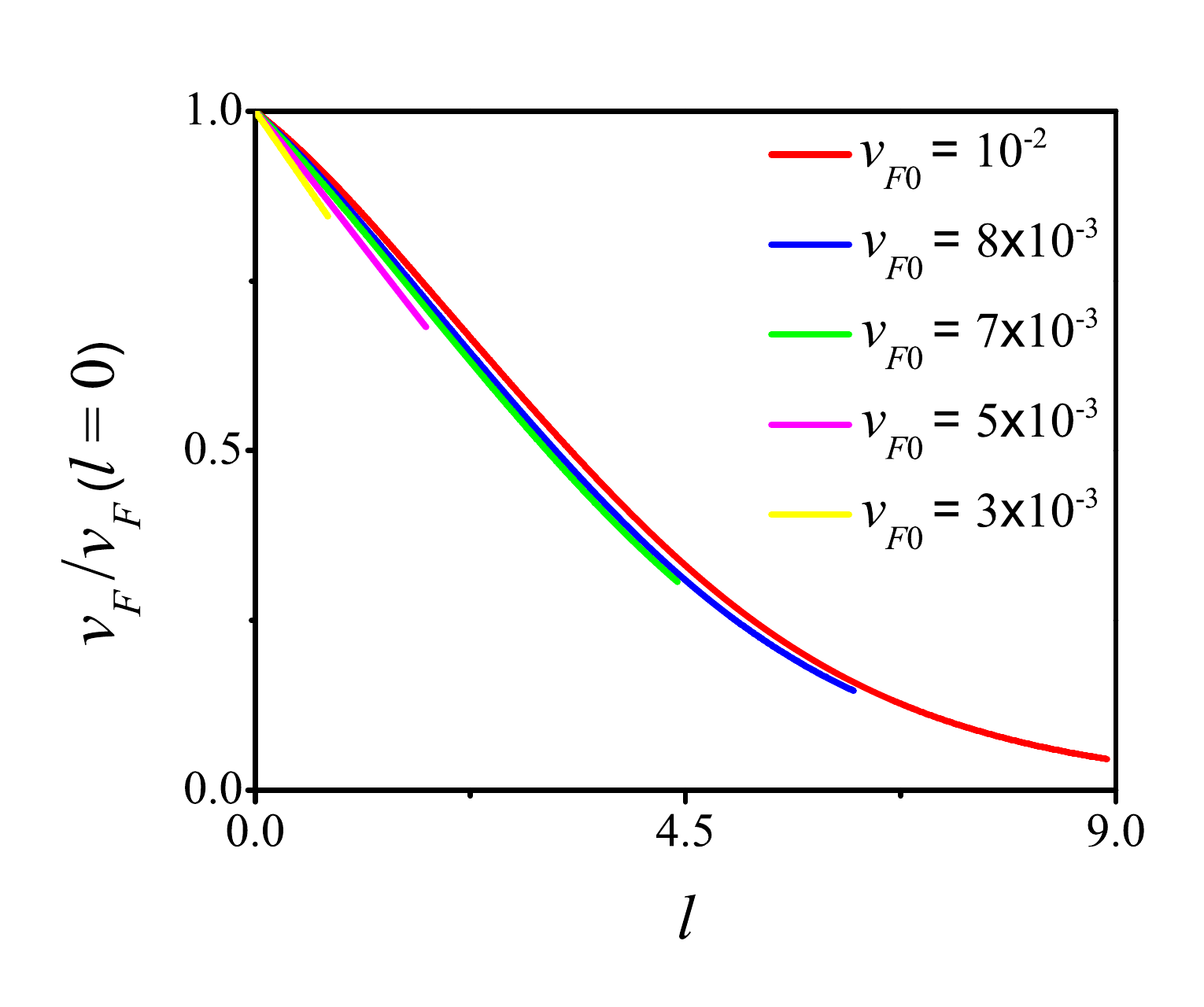}}
    \hspace{-0.3cm}
    \subfigure[ ]{\includegraphics[scale=0.21]{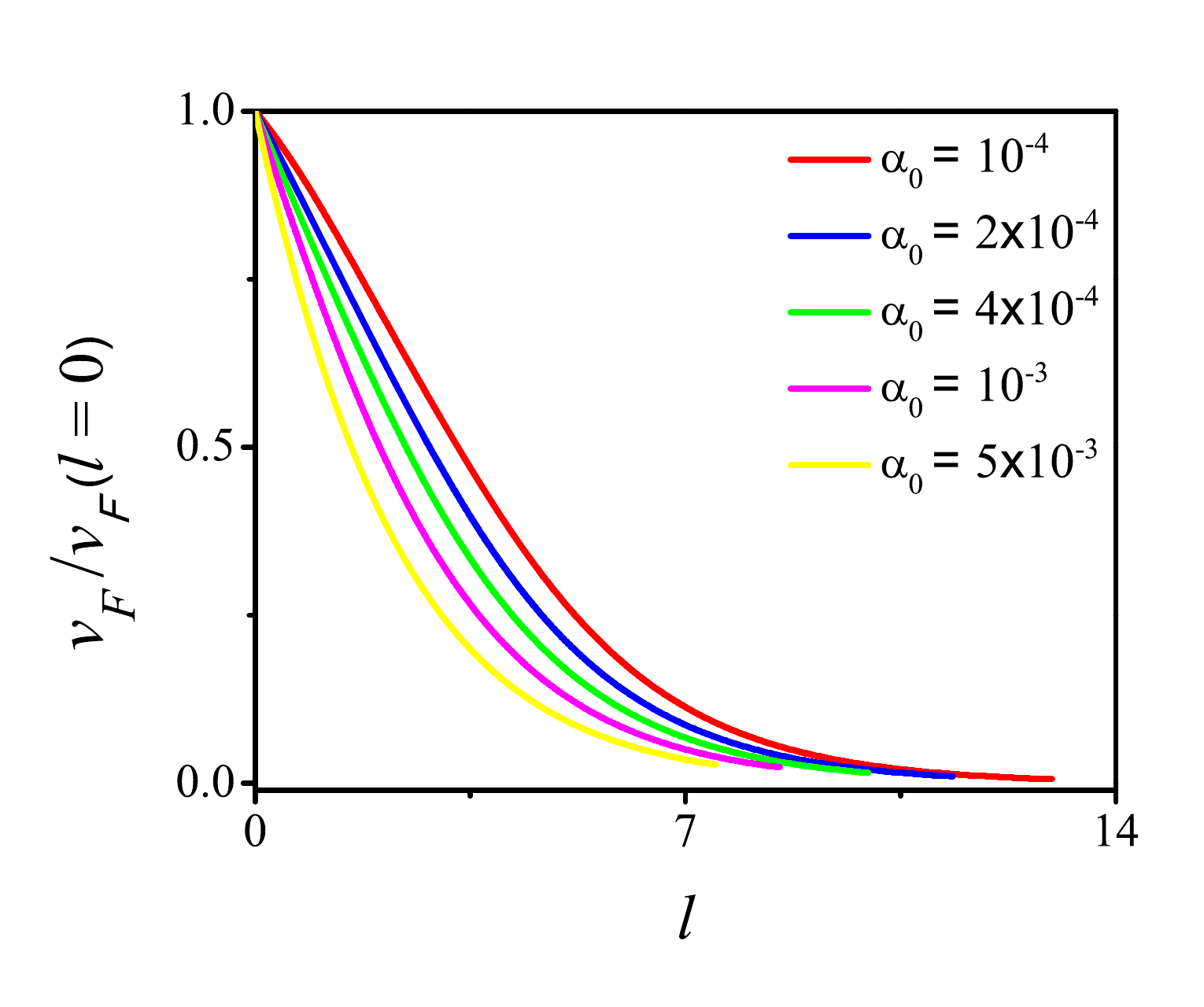}}
    \caption{(Color online) Energy-dependent evolutions of $v^{}_{F}/v^{}_{F} (l = 0)$ for both two strategies:
(a) $\zeta = 0.5$, $v_{F0} = 10^{-3}$, $\alpha_{0} = 10^{-4}$,
and $\lambda_{i0} = 10^{-5}$ (Strategy \uppercase\expandafter{\romannumeral 1}),
(b) Various sets of $\zeta$, $v_{F0}$, $\alpha_{0}$, and $\lambda_{i0}$ (Strategy \uppercase\expandafter{\romannumeral 1}),
(c) $\zeta$ with $v_{F0} = 10^{-4}$, $\alpha_{0} = 10^{-3}$,
and $\lambda_{i0} = 10^{-5}$ (Strategy \uppercase\expandafter{\romannumeral 2}),
(d) $\lambda_{i0}$ with $\zeta = 0.5$, $v_{F0} = 10^{-3}$,
and $\alpha_{0} = 10^{-3}$ (Strategy \uppercase\expandafter{\romannumeral 2}),
(e) $v_{F0}$ with $\zeta = 0.5$, $\alpha_{0} = 10^{-3}$,
and $\lambda_{i0} = 10^{-3}$ (Strategy \uppercase\expandafter{\romannumeral 2}),
and (f) $\alpha_{0}$ with $\zeta = 0.5$, $v_{F0} = 10^{-3}$,
and $\lambda_{i0} = 10^{-5}$ (Strategy \uppercase\expandafter{\romannumeral 2}). }
    \label{fig6}
\end{figure*}

Subsequently, going beyond Strategy \uppercase\expandafter{\romannumeral 1}, we now aim to develop an improved approach to handle the parameter $\eta$, dubbed Strategy \uppercase\expandafter{\romannumeral 2}.
Learning from RG Eqs. \eqref{RG_Eq.1} and \eqref{RG_Eq.2}, we notice that both $\alpha$ and $v_{F}$ are energy-dependent and
coupled with all the other interaction parameters. In addition, the variable $\eta$ as designated in
Sec.~\ref{Sec_RGEqs} measures the feature of energy dispersion and closely
hinges upon the relationship between $\alpha$ and $v_{F}$. Consequently, $\eta$ would be
influenced by the competition between these interaction parameters after taking into
account the effect of fluctuations. In this sense, it is necessary to determine the boundary
conditions of $\eta$ for further consideration.
To this end, we at first determine the boundary conditions for $\eta$, which is governed by
the energy $E$ that consists of $\zeta$, $\alpha$, $v_{F}$, $k_{x}$, and $k_{y}$.
With restricting $\alpha=v_{F}$, it becomes evident that $\eta$ is insensitive to the tilting parameter $\zeta$
as shown in Fig.~\ref{fig.isenergetic.profile}, and this accordingly suggests that $\eta$ reduces
to a momentum-dependent variable, namely $\eta(k_{x}, k_{y})$. Besides, the dependence of $\eta$ on $k_{x}$ and $k_{y}$ is equivalent to that of the angle $\varphi \equiv \arctan(k_{y}/k_{x})$ with $\left(-\frac{\pi}{2}<\varphi<\frac{\pi}{2}\right)$.
As a consequence, the general form of $\eta$ becomes $\eta(\alpha, v_{F}, \varphi)$, which is subject to the following restrictions due to the unique energy dispersion of a 2D tilted SDSM
\begin{align}
\eta\left \{
\begin{array} {l l}
\approx-0.5, \hspace{1.97cm} \alpha \gg v^{}_{F}, \\ \\
\approx-0.5, \hspace{1.97cm} \alpha \ll v^{}_{F}, \\ \\
=0.5\left|\tan \frac{\varphi}{2}\right|-0.5,  \hspace{0.25cm} \alpha=v^{}_{F}.
\end{array} \right.\label{Eq_eta_three-cases}
\end{align}
Based upon these boundary conditions, we present a trial expression for this very
parameter $\eta$ as follows,
\begin{align}
\eta(\alpha, v^{}_{F}, \varphi) & =\frac{1}{2} \left(\frac{\lg \left(1+\frac{4 \alpha v^{}_{F}}{\alpha^{2}+v^{2}_{F}}\right)}{\lg \left(2+\frac{2 \alpha}{\alpha+v^{}_{F}}\right)}\right)   \left|\tan \frac{\varphi}{2}\right| \notag \\
& ~\quad -\frac{1}{2} \left(\frac{\lg \left(1+\frac{4 \alpha}{\alpha+v^{}_{F}}\right)}{\lg \left(2+\frac{3 \alpha}{\alpha+2 v^{}_{F}}\right)}\right), \label{eta.Tri}
\end{align}
which yields after adopting the expectation value with respect to the directions of momenta,
\begin{align}
\eta(\alpha, v^{}_{F}) & =\frac{1}{\pi} \int_{-\frac{\pi}{2}}^{\frac{\pi}{2}} \mathrm{d} \varphi \, \eta(\alpha, v^{}_{F}, \varphi)  \notag \\
& =0.159  \left[1.386  \left(\frac{\lg \left(1+\frac{4 \alpha v^{}_{F}}{\alpha^{2}+v^{2}_{F}}\right)}{\lg \left(2+\frac{2 \alpha}{\alpha+v^{}_{F}}\right)}\right) \right. \notag \\
& ~\quad - \left. \pi  \left(\frac{\lg \left(1+\frac{4 \alpha}{\alpha+v^{}_{F}}\right)}{\lg \left(2+\frac{3 \alpha}{\alpha+2 v^{}_{F}}\right)}\right)\right] \label{eta.K1}.
\end{align}
In contrast to the constant $\eta$ in Strategy \uppercase\expandafter{\romannumeral 1}, the new parameter $\eta(\alpha, v_{F})$ in Strategy \uppercase\expandafter{\romannumeral 2} depends on the interaction parameters and, as a result, becomes energy-dependent. This energy-dependent parameter will be employed to elucidate the low-energy behavior of a 2D tilted SDSM in Sec.~\ref{Sec_phyiscal_behavior}.

\section{Low-energy critical behavior}\label{Sec_phyiscal_behavior}

There exist two distinct strategies for the parameter $\eta$ explained in Sec.~\ref{Sec_two_strategies}.
In Strategy \uppercase\expandafter{\romannumeral 1}, the parameter $\eta$ is identified as a constant with $\eta=-0.25$, whereas in Strategy \uppercase\expandafter{\romannumeral 2}, it is dependent upon the microstructural parameter and the Fermi velocity
as expressed in Eq.~(\ref{eta.K1}). We within this section are going to address the low-energy critical behavior of the interaction parameters for both such two strategies based on the RG equations~(\ref{RG_Eq.1})-(\ref{RG_Eq.6}).  It is of particular importance to highlight that Strategy I represents the simplest approximation for the $\eta$, namely,
considering a constant $\eta$, but instead Strategy II offers a well improved representation, where
we treat $\eta$ as a function of $\alpha$ and $v_{F}$. Under this circumstance, these two strategies may yield
different results while the initial conditions are suitable.

\subsection{Behavior of the strengths of fermion-fermions interactions}\label{Subsec_lambda}

At first, we examine the energy-dependent behavior of fermion-fermion couplings, which
are governed by coupled RG equations~(\ref{RG_Eq.1})-(\ref{RG_Eq.6}). Performing the detailed numerical analysis gives rise to the
basic tendencies of fermion-fermion interactions as presented in Figs.~\ref{fig2}-\ref{fig4} for both
Strategy I and Strategy II. It is necessary to emphasize that
the vertical axis of these figures is designated by
$\lambda_{i} (l) / \lambda_{i} (l=0)$. Hereby, the initial values of the fermion-fermion interactions themselves,
$\lambda^{}_{i} (l=0)$, are quite small, which are assigned from $10^{-5}$ to $10^{-3}$.

We begin with the Strategy I. Our numerical analysis suggests that whether fermion-fermion interactions diverge in the low-energy regime is closely dependent on the initial conditions. When the initial values of the fermion-fermion interactions and
the tilting parameter are small, Fig.~\ref{fig2}(a) with $\zeta=0.3$ shows that fermionic couplings decrease as the energy scale
decreases (or $l$ increases), eventually vanishing in the low-energy regime. Besides, these qualitative results are insensitive to the variations of both the fermion velocity ($v_{F0}$) and the microstructural parameter ($\alpha_0$).
In sharp contrast, while the initial values of fermionic couplings are taken big enough, the
fermion-fermion interactions ($\lambda_i$ with $i=0,1,2,3$) as illustrated in Fig.~\ref{fig2}(b)
with $\zeta=0.5$ quickly increase and then finally diverge at certain critical energy scale
denoted by $l_c$.  Hereby, it is of particular importance to emphasize that the critical energy scale $l_c$
is closely associated with the initial values of parameters $\zeta$, $v_F$, $\alpha$, and $\lambda_i$.
Let us take $\lambda_0$ for example (the basic tendencies for $\lambda_{1,2,3}$ are analogous).
In particular, a bigger tilting parameter $\zeta$ is helpful to the divergence of fermionic couplings as shown in Fig.~\ref{fig3}(a). Additionally, Fig.~\ref{fig4}(a)-(c) indicate a smaller $v_{F0}$ and $\alpha_0$ as well as a bigger $\lambda_{i0}$ can enhance the divergence and diminish the $l_c$.

Next, let us move to consider the Strategy II.
Compared to the Strategy I, we realize that fermion-fermion interactions in the Strategy II
can diverge at a critical energy $l_c$ much more easily. As illustrated in Fig.~\ref{fig2}(c),
they are divergent as approaching $l_c$, which are insensitive to the initial values of the tilting parameter
or other parameters. Although the divergent tendencies of fermionic couplings are free of the initial conditions,
the energy scale denoted by $l_c$ can be influenced by the related parameters.
Concretely, the bigger $\zeta$ gives rise to a smaller $l_c$  as shown in Fig.~\ref{fig3}(b).
In addition, Figs.~\ref{fig4}(d), \ref{fig4}(e), and \ref{fig4}(f) present that
smaller values of $v_{F0}$ and bigger initial values of $\lambda_{i}$ yield a smaller $l_c$.
These are qualitatively consistent with the influence of parameters on $l_c$ in the Strategy I.
However, in contrast to Strategy I, we find that a bigger value of the microstructural parameter $\alpha$ leads to
a bigger critical energy (or smaller $l_c$). 

\begin{figure*}[htbp]
\centering
\subfigure[ ]{\includegraphics[scale=0.21]{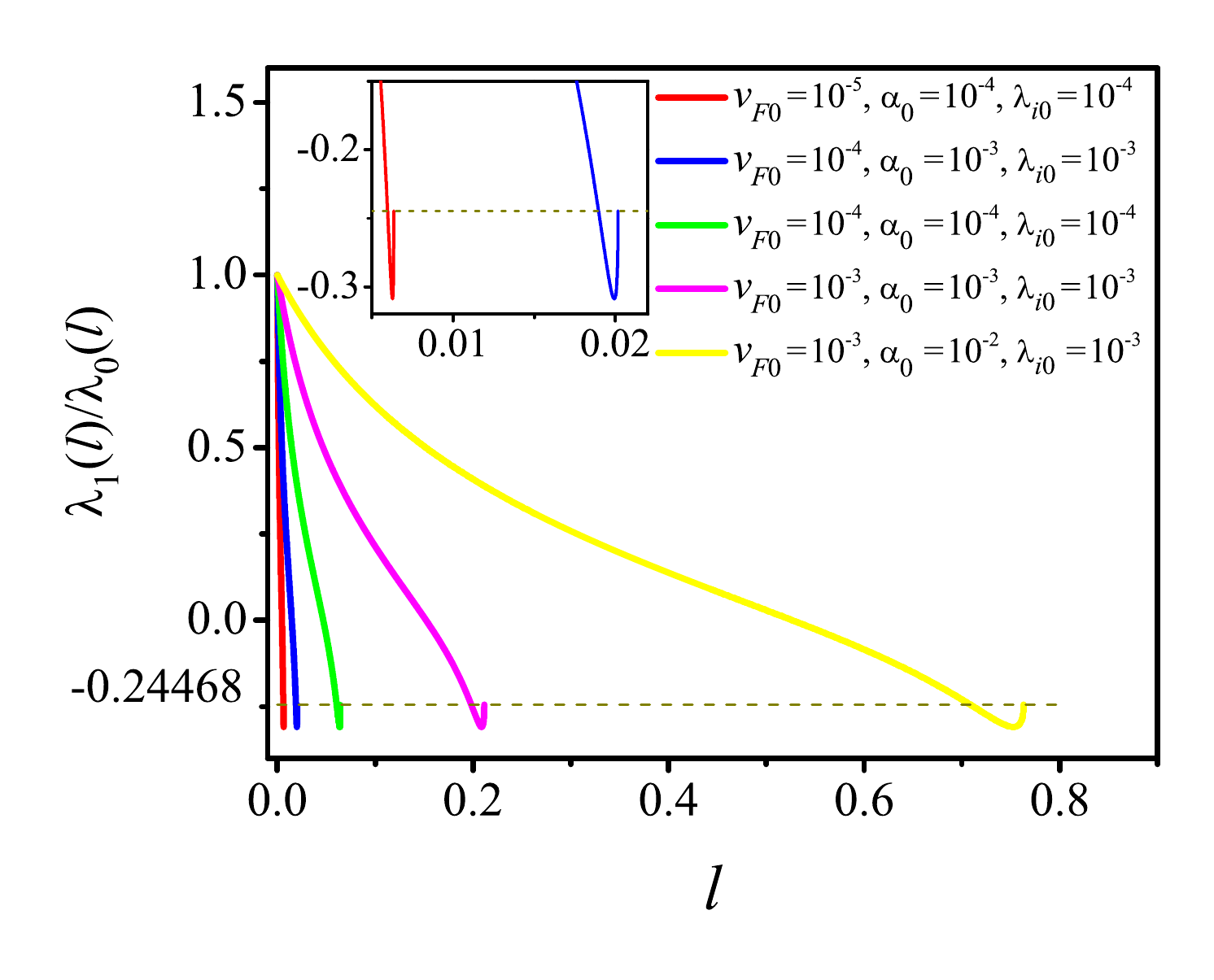}}
\hspace{-0.3cm}
\subfigure[ ]{\includegraphics[scale=0.21]{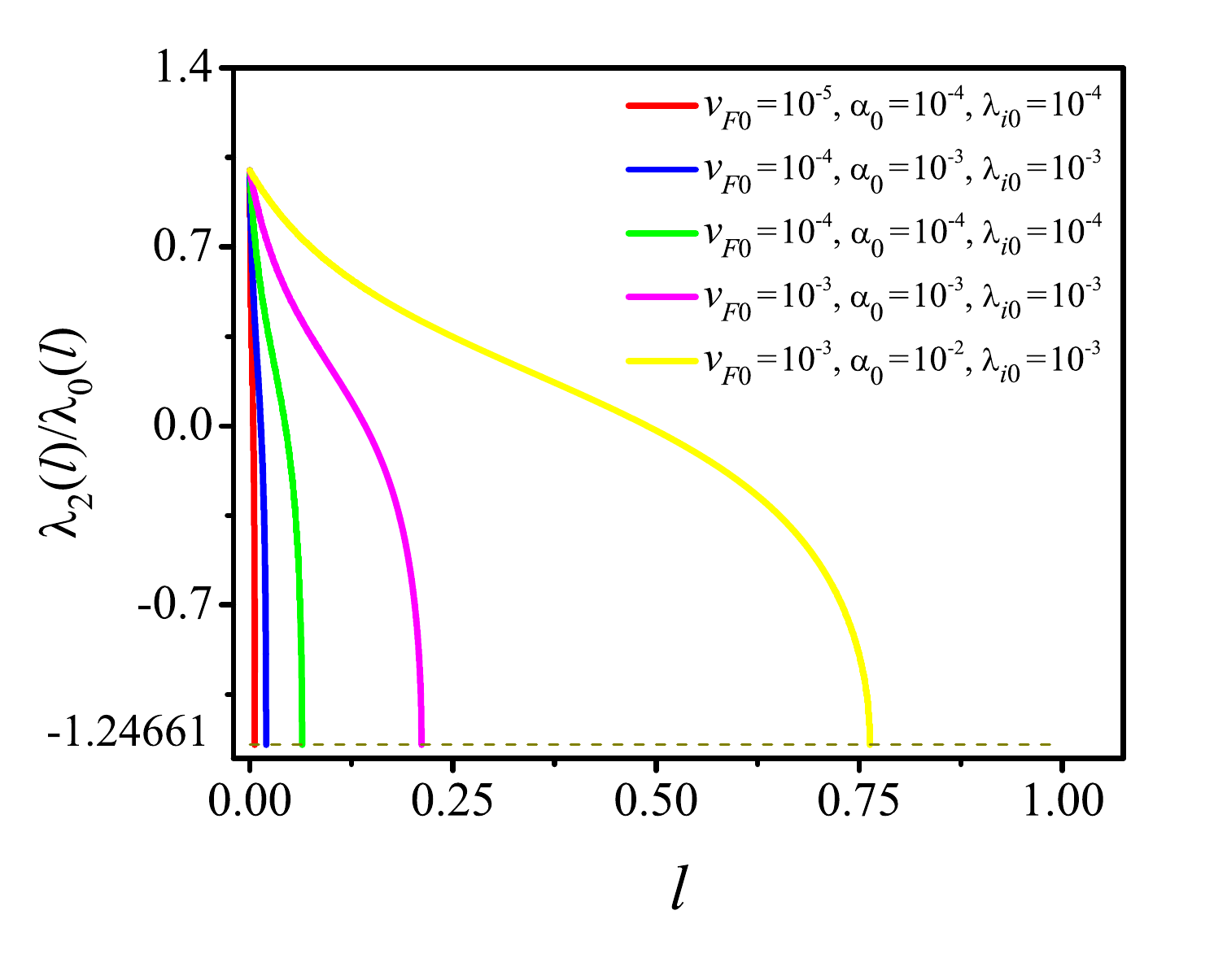}}
\hspace{-0.3cm}
\subfigure[ ]{\includegraphics[scale=0.21]{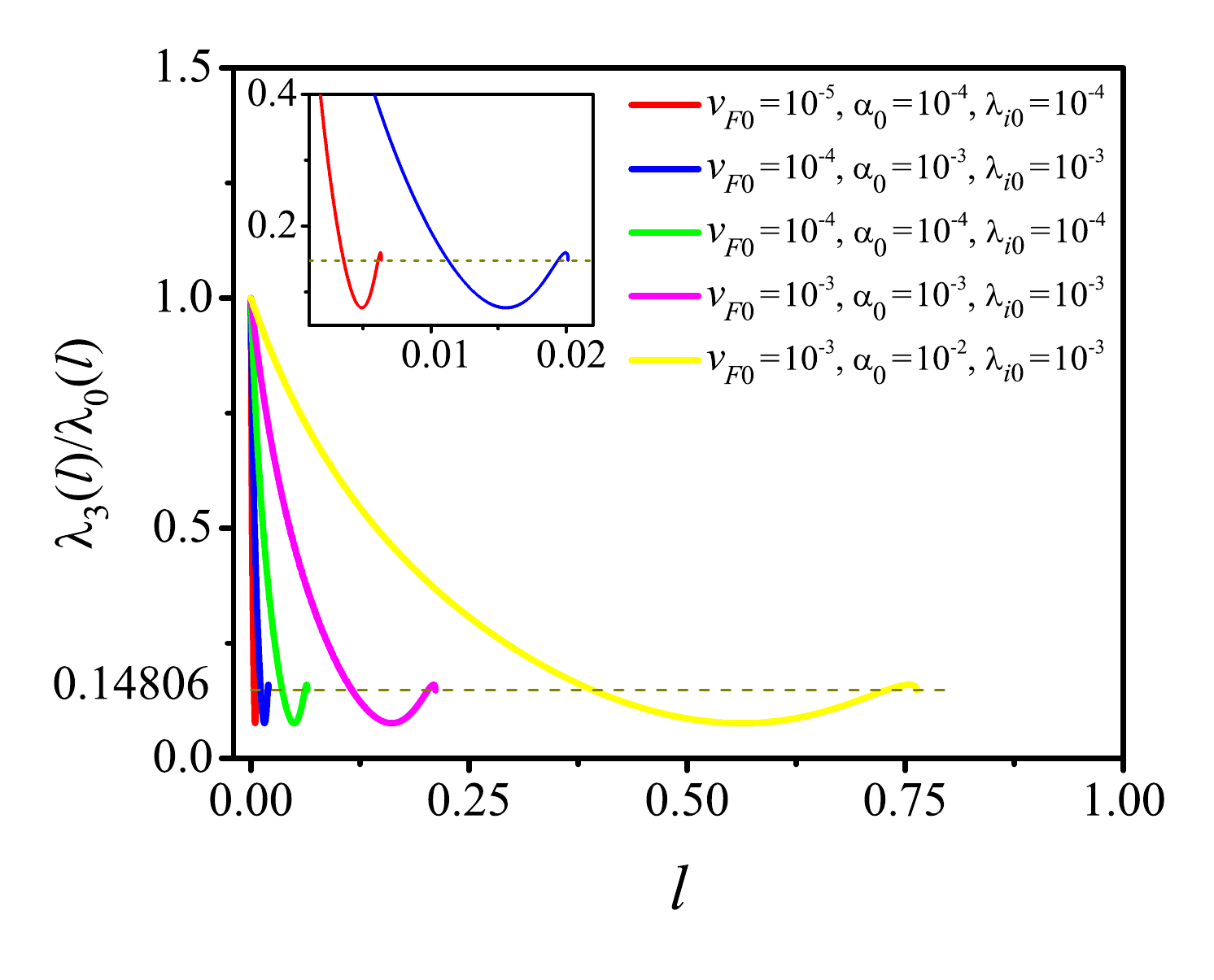}}\\\vspace{-0.3cm}
\subfigure[ ]{\includegraphics[scale=0.21]{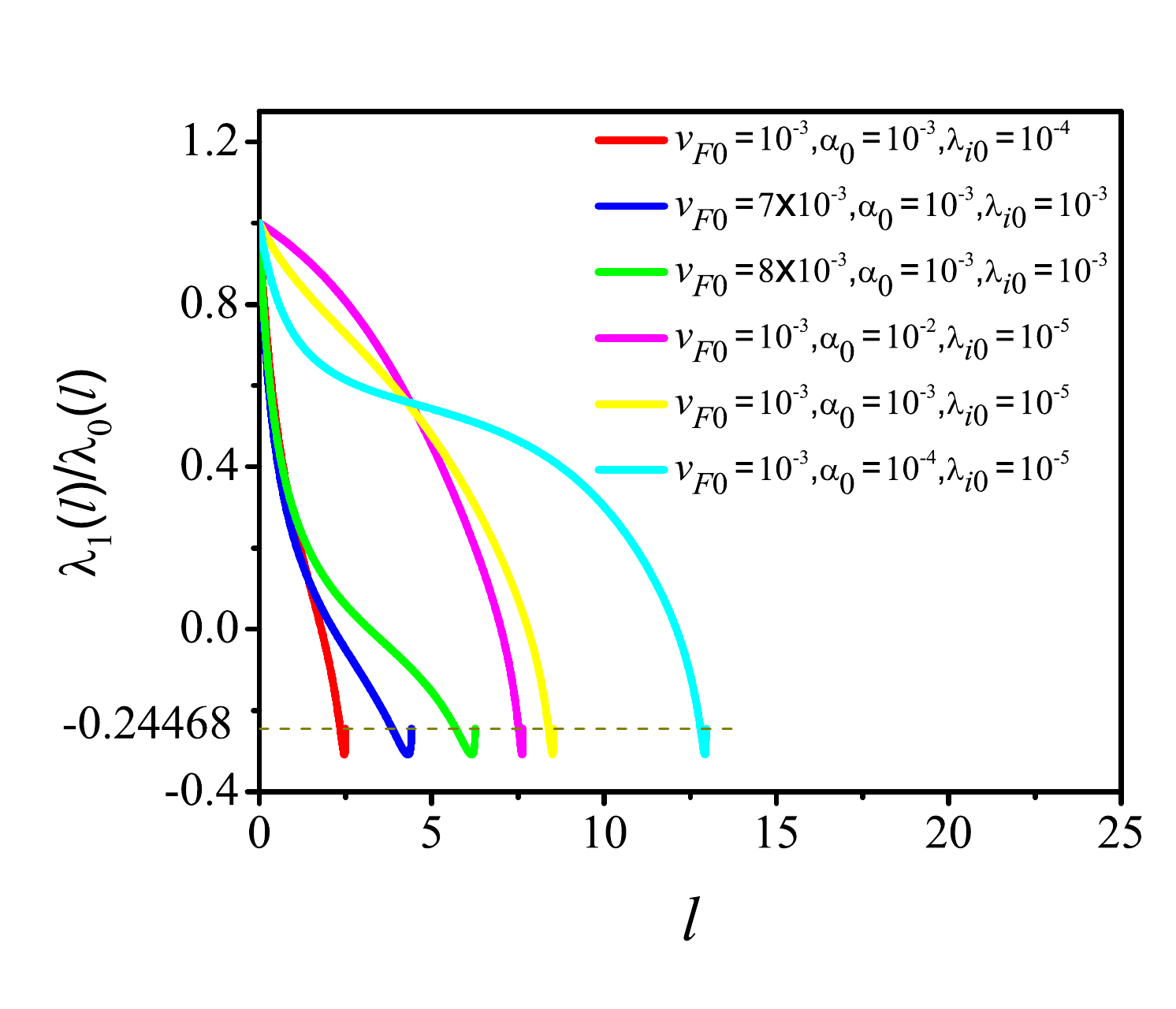}}
\hspace{-0.3cm}
\subfigure[ ]{\includegraphics[scale=0.21]{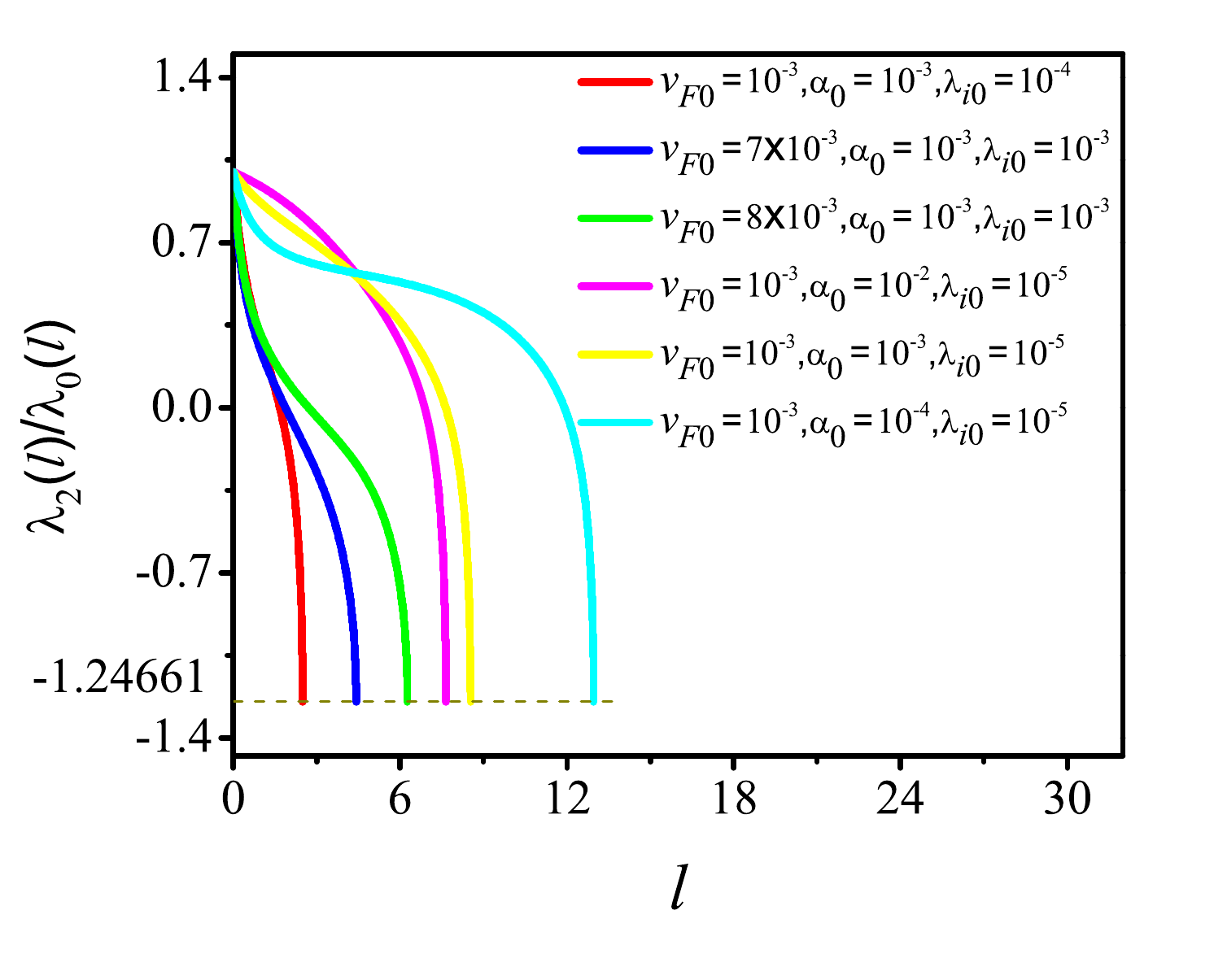}}
\hspace{-0.3cm}
\subfigure[ ]{\includegraphics[scale=0.21]{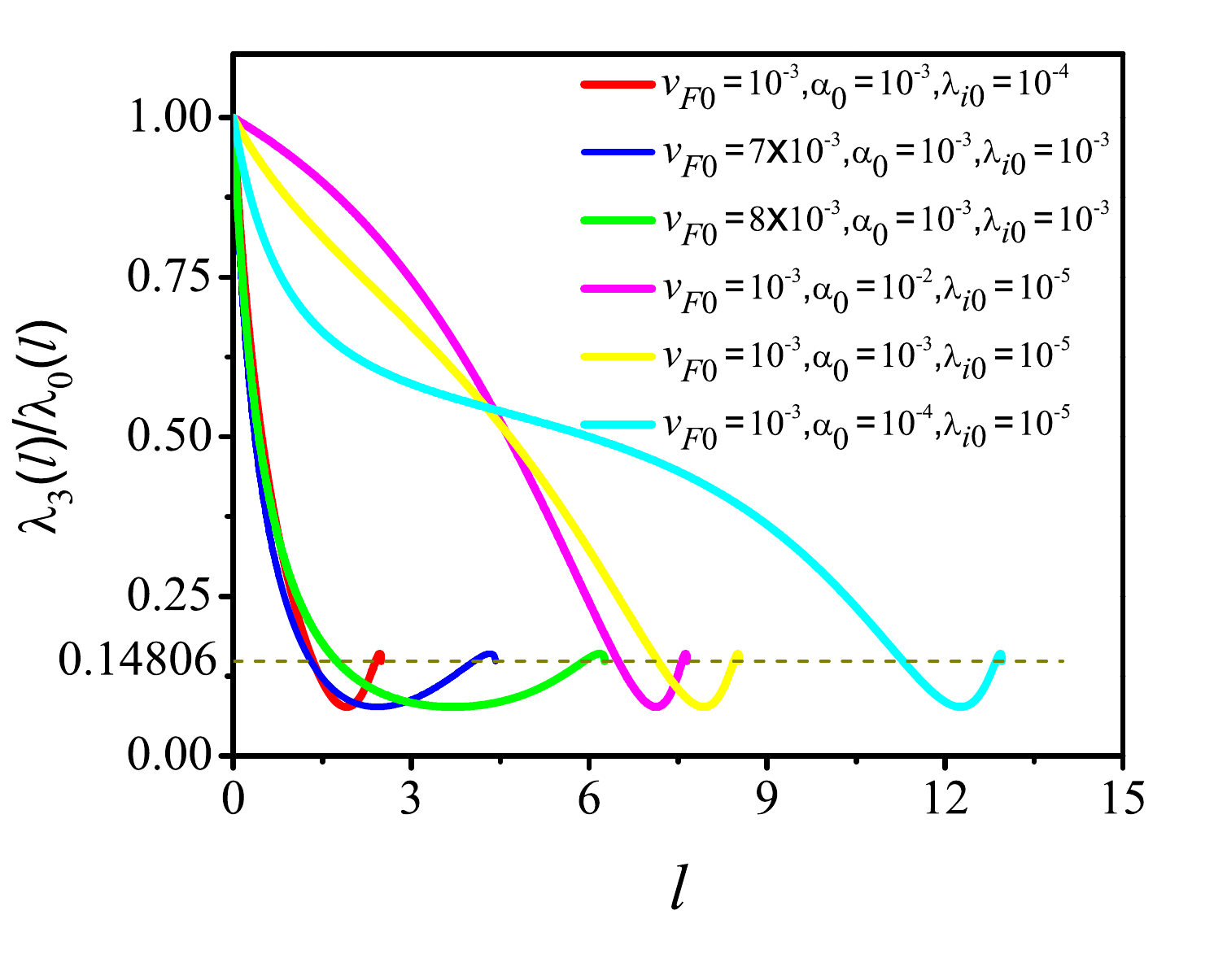}}
\caption{(Color online) Energy-dependent evolutions of $\lambda_j(l)/\lambda_0(l)$ ($j=1,2,3$) for both two Strategies:~(a) $\lambda_1(l)/\lambda_0(l)$, (b) $\lambda_2(l)/\lambda_0(l)$ and
(c) $\lambda_3(l)/\lambda_0(l)$ for (Strategy I), and (d) $\lambda_1(l)/\lambda_0(l)$, (e) $\lambda_2(l)/\lambda_0(l)$ and (f) $\lambda_3(l)/\lambda_0(l)$ for (Strategy II) with $\zeta = 0.5$. }
\label{fig7}
\end{figure*}

To be brief, for the Strategy I, whether fermion-fermion interactions $\lambda_i$ diverge in the low-energy region largely
depends on their initial values and the magnitude of the tilting parameter $\zeta$.
However, for the Strategy II, the fermion-fermion interactions usually flow divergently around a critical energy scale denoted
by $l_c$ in the low-energy regime.


\subsection{Fate of the microstructural parameter $\alpha$ and the Fermi velocity $v^{}_{F}$}\label{Subsec_alpha}

Subsequently, we shift our attention to the behavior of the microstructural parameter and the
Fermi velocity as the energy scale decreases.
For simplicity, the $v^{}_{F0}$, $\alpha^{}_{0}$, and $\lambda^{}_{i0}$ are utilized to serve as
the initial values of the related parameters at $l=0$. Fig.~\ref{fig5} and Fig.~\ref{fig6} provide the primary results of
$\alpha$ and $v_F$ for Strategy \uppercase\expandafter{\romannumeral 1} and Strategy \uppercase\expandafter{\romannumeral 2},
respectively.

Specifically, let us begin with the Strategy \uppercase\expandafter{\romannumeral 1}.
On one side, if the fermion-fermion couplings are not divergent as shown in Fig.~\ref{fig5}(a) and Fig.~\ref{fig6}(a)
with the small initial values and tilting parameter,
$\alpha/\alpha^{}_{0}$ monotonously increases, while $v^{}_{F}/v^{}_{F0}$ gradually decreases
and vanishes with reducing the energy scale.
On the other side, once the fermion-fermion interactions, armed with the bigger $\lambda_{i0}$ and $\zeta$,
go toward divergence at a critical energy scale denoted by $l^{}_{c}$,
the RG equations are required to stop at $l=l_c$. This implies that the $\alpha$ increases and reaches its
maximum achievable value $\alpha (l^{}_{c})$ as depicted in Fig.~\ref{fig5}(b). Meanwhile,
as illustrated in Fig.~\ref{fig6}(b), the Fermi velocity falls until it arrives at a finite value, i.e., $v^{}_{F} (l^{}_{c})$.
Besides, the basic evolutions of both $\alpha/\alpha^{}_{0}$ and $v^{}_{F}/v^{}_{F0}$ are insensitive to
the tilting parameter, $v^{}_{F0}$, $\alpha^{}_{0}$, and $\lambda^{}_{i0}$.

\begin{table}[htbp]
\caption{The qualitative impacts of the absolute value of $\zeta$ and the initial values
of $\lambda^{}_{i}$, $\alpha$, and $v^{}_{F}$ on the critical energy scale $l^{}_{c}$, and the
corresponding critical values $\alpha(l^{}_{c})$ and $v^{}_{F} (l^{}_{c})$ for the situations with
the divergent fermion-fermion interactions. Hereby, ``$+$'' and ``$-$'' are adopted to indicate the rise and drop of the critical values.
In addition, ``$\uparrow$'', ``$\downarrow$'', and ``$\star$'' stand for increasing, decreasing, and fixing the values of the related parameters, respectively.}
\vspace{0.4cm}
\centering
\begin{tabular}{c    |c   c  c    c |   c c c}
\hline \hline
\specialrule{0em}{1pt}{1pt}
  &   $\lambda^{}_{i0}$  &  $\alpha^{}_{0} $    &  $v^{}_{F0} $   &  $|\zeta|$     & $\alpha(l^{}_{c})$ & $v^{}_{F} (l^{}_{c})$ & \hspace{9pt}$l^{}_{c}$ \hspace{9pt}\\
\hline
\specialrule{0em}{1pt}{1pt}
\multirow{4}{*}{Strategy \uppercase\expandafter{\romannumeral 1}}
&  \hspace{2pt}$\uparrow$ \hspace{2pt}& \hspace{2pt}$\star$ \hspace{2pt} & \hspace{2pt}$\star$\hspace{2pt}   & \hspace{2pt}$\star$  \hspace{2pt}& $-$ &$+$ & $-$\\
& $\star$ & $\uparrow$ & $\star$   &  $\star$  &  $+$ & $-$&  $+$ \\
&  $\star$ & $\star$ & $\downarrow$   & $\star$ & $-$ & $+$ &  $-$  \\
& $\star$  & $\star$  & $\star$  & $\uparrow$  &  $-$ & $+$ & $-$   \\
\hline
\specialrule{0em}{1pt}{1pt}
\multirow{4}{*}{\hspace{1pt} Strategy \uppercase\expandafter{\romannumeral 2}} \hspace{1pt}
&   $\uparrow$ & $\star$  & $\star$   & $\star$  &  $-$ & $+$ & $-$  \\
&   $\star$ & $\downarrow$ & $\star$   &  $\star$   & $+$ & $-$ & $+$  \\
&    $\star$ & $\star$ & $\downarrow$   & $\star$  & $-$ & $+$ & $-$  \\
&  $\star$  & $\star$  & $\star$  & $\uparrow$  &  $-$ & $+$ & $-$   \\
\hline \hline
\end{tabular}
\label{Tab1}
\end{table}

To proceed, we consider the Strategy \uppercase\expandafter{\romannumeral 2}.
Learning from Fig.~\ref{fig5}(c)-(f) and Fig.~\ref{fig6}(c)-(f), we notice that the basic properties
of both $\alpha/\alpha^{}_{0}$ and $v^{}_{F}/v^{}_{F0}$ analogous to their Strategy \uppercase\expandafter{\romannumeral 1}
counterparts. To be specific, Fig.~\ref{fig5}(c)-(d) and Fig.~\ref{fig6}(c)-(d) suggest that the tendencies of $\alpha$ or $v^{}_{F}$ in Strategy \uppercase\expandafter{\romannumeral 2} are insusceptible to the tilting parameter. In contrast to the Strategy \uppercase\expandafter{\romannumeral 1}, $v_{F0}$ and $\alpha_{0}$ are capable of considerably
reshaping the slopes of $\alpha/\alpha_{0}$ or $v_{F}/v_{F0}$ as illustrated in Fig.~\ref{fig5}(e)-(f) and
Fig.~\ref{fig6}(e)-(f). Moreover, we realize that the critical energy scales are closely associated with
the initial values of interaction parameters. In particular, Fig.~\ref{fig5}(e)-(f) and Fig.~\ref{fig6}(e)-(f)
indicate that a smaller $\alpha_{0}$ yields a lower critical energy scale (a bigger $l_c$), but instead a smaller initial value of
Fermi velocity gives rise to a higher critical energy scale (a smaller $l_c$).

For convenience, we present Table~\ref{Tab1} to summarize the key points for the behavior of $\alpha$ and $v_F$ as well as
the critical energy scale. One can find that the variations of $\zeta$, $v^{}_{F0}$, or $\lambda^{}_{i0}$ bring the similar consequences on the critical energy scale and its accompanied critical values ($v^{}_{F} (l^{}_{c})$, $\alpha (l^{}_{c})$) for both two strategies.
In comparison, a bigger $\alpha^{}_{0}$ leads to a bigger $\alpha (l^{}_{c})$ in Strategy \uppercase\expandafter{\romannumeral 1}, which differs from Strategy \uppercase\expandafter{\romannumeral 2}. This difference can be understood as follows.
In Strategy \uppercase\expandafter{\romannumeral 1}, $\eta$ remains a constant and thus cannot
influence the fermion-fermion interactions.
However, the variable $\eta$~(\ref{eta.K1}) is dependent on $\alpha$ and $v_F$ for the
Strategy \uppercase\expandafter{\romannumeral 2}, which are
coupled with the fermionic interactions through the RG equations.

\begin{figure}[htbp]
\centering
\includegraphics[scale=0.25]{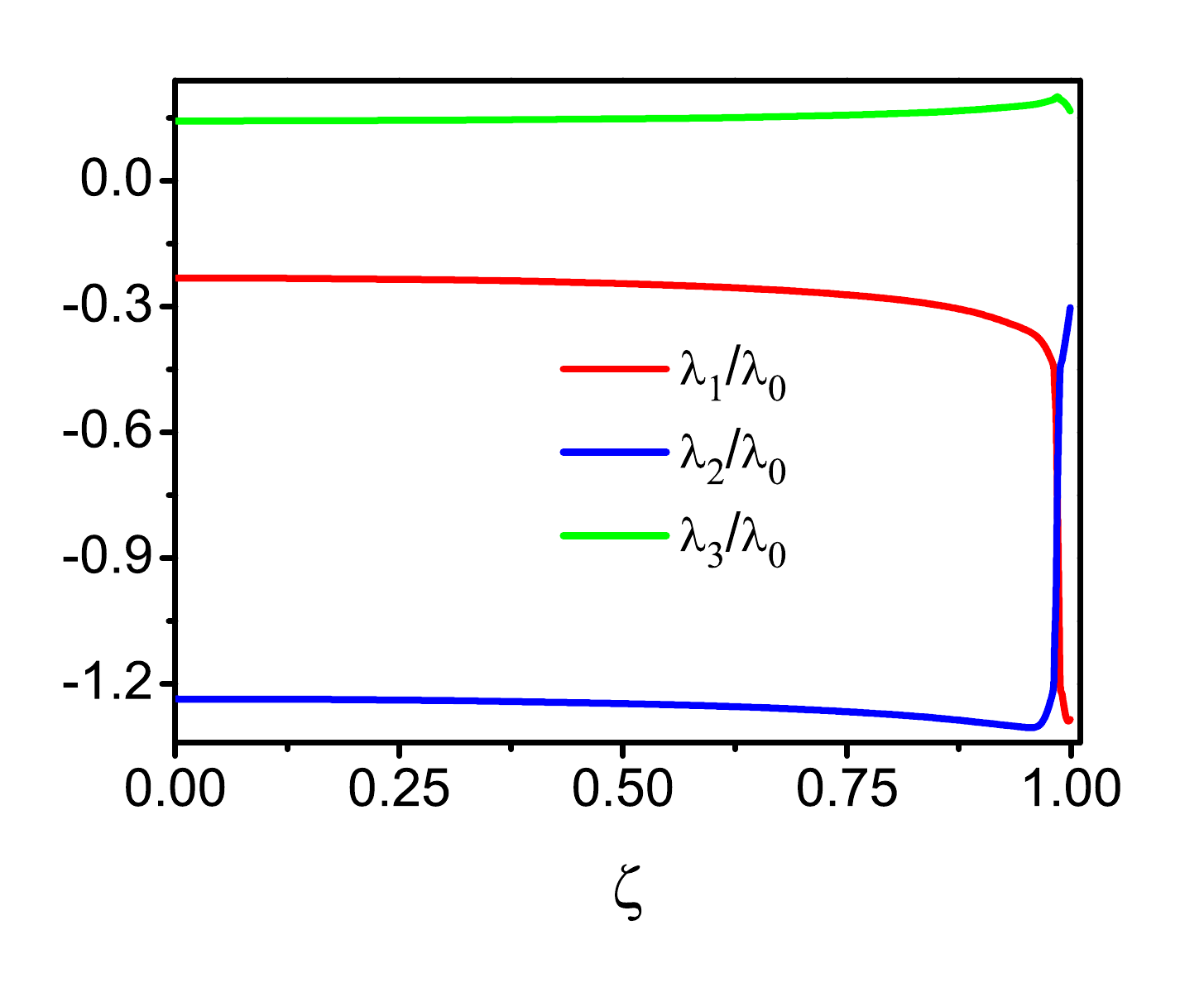}
\vspace{-0.39cm}
\caption{(Color online) $\zeta$-dependent evolutions of the RFP (i.e., $\lambda_j(l_c)/\lambda_0(l_c)$) with $v_{F0} = 10^{-4}$, $\alpha_{0} =  10^{-3}$, and $\lambda_{i0} =  10^{-4}$ for the Strategy II (the basic tendencies for the situations
with the divergent fermion-fermion interactions in Strategy I are analogous and not shown hereby.}
\label{fig8}
\end{figure}

\subsection{Behavior of the relatively fixed point}\label{Subsec_RFP}

At last, we put our focus on the stability of the relatively fixed point (RFP). Hereby,
the RFP describes the behavior of the interaction parameters at a critical energy scale, $l_c$,
beyond which the RG equations become invalid~\cite{Murray2014PRB,Roy2018PRX2,DZZW2020PRB,Chubukov2016PRX,Chubukov2010,Nandkishore2012NP}.
Specifically, we hereby define the coordinates of the RFP by employing the scalings
$\lambda_j(l_c)/\lambda_0(l_c)$ with $j=0,1,2,3$.

In order to determine the
RFP, we need to examine the flows of $\lambda_j(l)/\lambda_0(l)$, where $\lambda_0(l)$
evolves with decreasing energy without changing sign.
After performing the careful numerical analysis as shown in Fig.~\ref{fig7}, we find that
all relevant parameters can only quantitatively modify the tendencies of $\lambda_j(l)/\lambda_0(l)$ but
the final values of $\lambda_j(l_c)/\lambda_0(l_c)$ near the RFP approximately
arrive at \textbf{$(-0.24468, -1.24661, 0.14806)$}, which
are stable under the variations of the initial conditions.
For the Strategy I, the bigger initial values of the fermion-fermion interactions $\lambda_{i}$,
and smaller values of $v_{F0}$ as well as $\alpha_0$ are all favorable to increase the critical energy scale (or
reduce $l_c$). In contrast, for the Strategy II, the bigger values of $v_{F0}$ and $\lambda_{i0}$ increase the critical energy (i.e., smaller $l_c$), while the smaller values of $\alpha_0$ increase $l_c$. In sharp contrast, we notice that the RFP is highly sensitive to the tilting parameter $\zeta$, as depicted in Fig.~\ref{fig8} for representative initial parameter values.
To be concrete, $\lambda_1(l_c)/\lambda_0(l_c)$ and $\lambda_2(l_c)/\lambda_0(l_c)$ remain stable for $\zeta$ values ranging from 0 to 0.75, but, as $\zeta$ approaches 0.995, they decrease and increase sharply, respectively. For $\lambda_3(l_c)/\lambda_0(l_c)$, there exists a gradually increase over most of the $\zeta$ range, then followed by a rapid decrease beyond $\zeta=0.985$. These results are analogous for the Strategies I and II.

To wrap up, we schematically summarize the main conclusions related to the RFP in Fig.~\ref{fig9}. In both Strategy I and Strategy II, depending on the evolution of $\lambda_i$, the RFP emerges at $l_c$ where they diverge, while convergence occurs at the Gaussian fixed point (GFP). The behavior of the RFP is consistent with the conclusions of Sec.~\ref{Subsec_lambda} and is highly sensitive to the tilting parameter $\zeta$.


\section{Summary}\label{Sec_summary}

In summary, we utilize the RG approach~\cite{Wilson1975RMP,Polchinski1992arXiv,Shankar1994RMP}
to unbiasedly examine the influence of short-range
fermion-fermion interactions on the low-energy physical behavior in the 2D tilted SDSM.
After taking into account all the one-loop corrections, the coupled RG equations of all related parameters
are derived, which are expected to capture the low-energy physics of the tilted SDSM.
Due to the unique dispersion relation of the 2D tilted SDSM, we classify two strategies (i.e., Strategy I and II
designated in Sec.~\ref{Sec_two_strategies}) to make the analysis more convenient and clear.
A detailed numerical analysis is followed to present the low-energy properties of fermion-fermion interactions,
microstructural parameter, and fermion velocity, as well as the
behavior the RFP around the critical energy scales.

\begin{figure}[htbp]
    \centering
    \includegraphics[scale=1]{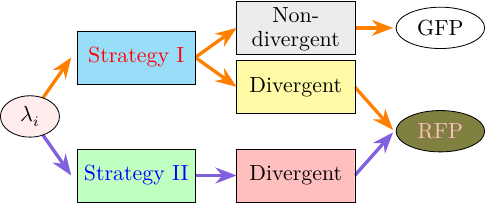}
    \caption{(Color online) Schematic illustration for the RFP and Gaussian fixed point (GFP)
    that is $(0, 0, 0, 0)$ and trivial in the low-energy regime.}
    \label{fig9}
\end{figure}

At first, we notice that the behavior of fermion-fermion interactions in the low-energy regime can either vanish or diverge. With respect to the Strategy I, the low-energy behavior of fermion-fermion interactions depends on both the interaction strength itself and the initial value of the tilting parameter as shown in Fig.~\ref{fig2}. For the weak initial values of the interactions and tilting parameter, the fermion-fermion couplings decrease and eventually vanish. In contrast, if the initial values of the related parameter are sufficiently large, they can increase and  diverge at $l=l_c$. However, as for the Strategy II, regardless of the initial values of the relevant parameters, the fermion-fermion couplings can diverge at a certain critical energy scale. In particular, the bigger tilting parameter, the microstructural parameter, and fermion-fermion interactions can increase the critical energy scale, corresponding to a smaller $l_c$, but instead the bigger fermion velocity prefers to decrease the critical energy scale.
Next, we examine the low-energy behavior of the microstructural parameter $\alpha$ and the fermionic velocity $v_F$.
It can be found that the fermion-fermion couplings drive an increase in $\alpha$ but a decrease in $v_F$, respectively.
As to the Strategy I, if the fermion-fermion interactions do not diverge, $\alpha$ increases and $v_F$ decreases gradually in the low-energy regime. In contrast, once the fermion-fermion interactions $\lambda_i$ diverge at the critical energy $l_c$
for either the Strategy I or the Strategy II, the microstructural parameter and fermion velocity would be truncated at $l=l_c$,
namely obtaining the final values $\alpha(l_c)$ and $v_F(l_c)$ as listed in Table.~\ref{Tab1} for distinct kinds of initial conditions. Furthermore, we find that, depending the fermion-fermion interactions vanish or diverge in the low-energy regime,
the system can either flow toward the GFP or RFP, which is defined by $\lambda_j(l_c)/\lambda_0(l_c)$,
as schematically shown in Fig.~\ref{fig9}. With a fixed tilting parameter, the RFP depicted in Fig.~\ref{fig7} is very stable against the variations of other parameters. However, Fig.~\ref{fig8} indicates that the RFP is sensitive to the initial value of tilting parameter $\zeta$ for both strategies. To recapitulate, we show that the fermion-fermion interactions are closely associated with the related parameters in the low-energy regime, which induce certain RFP at the critical energy scales. Future research is necessary to explore the observable quantities and phase transitions for
the 2D tilted SDSM and analogous materials. These may offer valuable insights and scientific significance for
future applications and related studies.

\section*{ACKNOWLEDGEMENTS}

W.L. has been supported by the Young Teacher Education and Research Project of Fujian Province,
China under Grant No. JAT200587) and thanks Yi-Sheng Fu for kind help in the numerical calculations.
X.Z.C. is very grateful to M.S. Jie-Qiong Li for helpful discussions.
J.W. was partially supported by the National Natural Science Foundation of China
under Grant No. 11504360.

\appendix

\section{One-loop corrections}\label{Appendix_1L-corrections}

\begin{figure}[htbp]
\centering
\includegraphics[scale = 1]{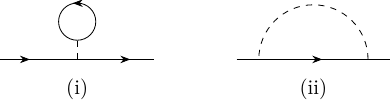}
\\
\vspace{-0.05cm}
\caption{One-loop corrections to the fermionic propagator due to the fermion-fermion interaction (the solid
and dash lines indicate the fermionic propagator and the fermion-fermion interaction, respectively).}
\label{fig10}
\end{figure}

The one-loop corrections to the fermionic propagator and fermion-fermion couplings
are depicted in Fig.~\ref{fig10} and Fig.~\ref{fig11}, respectively.
After long but straightforward calculations, we find that
Fig.~\ref{fig10} does not give rise to the nontrivial corrections to the fermionic propagator
(i.e, $\delta S_{\mathrm{fp}} = 0$). In comparison, Fig.~\ref{fig11} yields the one-loop corrections to fermion-fermion interactions
\begin{widetext}
\begin{align}
\delta S_{\mathrm{f}-\mathrm{f}}^{\lambda^{}_{0}}
& = \int_{-\infty}^{+\infty} \frac{d \omega^{}_{1} d \omega^{}_{2} d \omega^{}_{3}}{(2 \pi)^{3}} \int \frac{d^{2} \mathbf{k}^{}_{1} d^{2} \mathbf{k}^{}_{2} d^{2} \mathbf{k}^{}_{3}}{(2 \pi)^{6}}  \Psi^{\dagger}\left(\omega^{}_{1}, \mathbf{k}^{}_{1}\right) \sigma^{}_{0} \Psi\left(\omega^{}_{2}, \mathbf{k}^{}_{2}\right) \Psi^{\dagger}\left(\omega^{}_{3}, \mathbf{k}^{}_{3}\right) \sigma^{}_{0} \Psi\left(\omega^{}_{1}+\omega^{}_{2}-\omega^{}_{3}, \mathbf{k}^{}_{1}+\mathbf{k}^{}_{2}-\mathbf{k}^{}_{3}\right)  \notag \\
& \quad \times\frac{l}{4 \pi^{2} v^{}_{F}\sqrt{\alpha}}\left[\left(\lambda_{0}^{2}+\lambda_{1}^{2}+\lambda_{2}^{2}+\lambda_{3}^{2}\right) \mathcal{F}^{}_{1}  -2 \lambda_{0} \lambda_{1} \mathcal{F}^{}_{2} -2 \lambda_{0} \lambda_{2} \mathcal{F}^{}_{3}\right] ,\label{1LC.for.FF.0}  \\
\delta S_{\mathrm{f}-\mathrm{f}}^{\lambda_{1}}
& =\int_{-\infty}^{+\infty} \frac{d \omega_{1} d \omega_{2} d \omega_{3}}{(2 \pi)^{3}} \int \frac{d^{2} \mathbf{k}_{1} d^{2} \mathbf{k}_{2} d^{2} \mathbf{k}_{3}}{(2 \pi)^{6}}
\Psi^{\dagger}\left(\omega_{1}, \mathbf{k}_{1}\right) \sigma_{1} \Psi\left(\omega_{2}, \mathbf{k}_{2}\right) \Psi^{\dagger}\left(\omega_{3}, \mathbf{k}_{3}\right) \sigma_{1} \Psi\left(\omega_{1}+\omega_{2}-\omega_{3}, \mathbf{k}_{1}+\mathbf{k}_{2}-\mathbf{k}_{3}\right)\notag \\
& \quad \times\frac{l}{4 \pi^{2} v^{}_{F}\sqrt{\alpha}}\left[2 \lambda_{0} \lambda_{1} \mathcal{F}^{}_{1}  -\left(\lambda_{0}^{2}+\lambda_{1}^{2}+\lambda_{2}^{2}+\lambda_{3}^{2}\right) \mathcal{F}^{}_{2} -2\left(2 \lambda_{1}^{2}+\lambda_{0} \lambda_{1}+\lambda_{0} \lambda_{3}-\lambda_{1} \lambda_{2}-\lambda_{1} \lambda_{3}\right) \mathcal{F}^{}_{3}\right] , \label{1LC.for.FF.1} \\
\delta S_{\mathrm{f}-\mathrm{f}}^{\lambda_{2}}
& =\int_{-\infty}^{+\infty} \frac{d \omega_{1} d \omega_{2} d \omega_{3}}{(2 \pi)^{3}} \int \frac{d^{2} \mathbf{k}_{1} d^{2} \mathbf{k}_{2} d^{2} \mathbf{k}_{3}}{(2 \pi)^{6}}  \Psi^{\dagger}\left(\omega_{1}, \mathbf{k}_{1}\right) \sigma_{2} \Psi\left(\omega_{2}, \mathbf{k}_{2}\right) \Psi^{\dagger}\left(\omega_{3}, \mathbf{k}_{3}\right) \sigma_{2} \Psi\left(\omega_{1}+\omega_{2}-\omega_{3}, \mathbf{k}_{1}+\mathbf{k}_{2}-\mathbf{k}_{3}\right)  \notag \\
& \quad \times\frac{l}{4 \pi^{2} v^{}_{F}\sqrt{\alpha}}\left[2 \lambda_{0} \lambda_{2} \mathcal{F}^{}_{1} -2\left(2 \lambda_{2}^{2}+\lambda_{0} \lambda_{2}+\lambda_{0} \lambda_{3}-\lambda_{1} \lambda_{2}-\lambda_{2} \lambda_{3}\right) \mathcal{F}^{}_{2} -\left(\lambda_{0}^{2}+\lambda_{1}^{2}+\lambda_{2}^{2}+\lambda_{3}^{2}\right) \mathcal{F}^{}_{3}\right] , \label{1LC.for.FF.2}\\
\delta S_{\mathrm{f}-\mathrm{f}}^{\lambda_{3}}
& =\int_{-\infty}^{+\infty} \frac{d \omega_{1} d \omega_{2} d \omega_{3}}{(2 \pi)^{3}} \int \frac{d^{2} \mathbf{k}_{1} d^{2} \mathbf{k}_{2} d^{2} \mathbf{k}_{3}}{(2 \pi)^{6}}  \Psi^{\dagger}\left(\omega_{1}, \mathbf{k}_{1}\right) \sigma_{3} \Psi\left(\omega_{2}, \mathbf{k}_{2}\right) \Psi^{\dagger}\left(\omega_{3}, \mathbf{k}_{3}\right) \sigma_{3} \Psi\left(\omega_{1}+\omega_{2}-\omega_{3}, \mathbf{k}_{1}+\mathbf{k}_{2}-\mathbf{k}_{3}\right)  \notag \\
& \quad \times\frac{l}{4 \pi^{2} v^{}_{F}\sqrt{\alpha}}\left[2\left(-2 \lambda_{3}^{2}+\lambda_{1} \lambda_{3}+\lambda_{2} \lambda_{3}\right) \mathcal{F}^{}_{1} -2 \lambda_{0} \lambda_{2} \mathcal{F}^{}_{2} -2 \lambda_{0} \lambda_{1} \mathcal{F}^{}_{3}\right] .\label{1LC.for.FF.3}
\end{align}
\begin{figure*}[htbp]
\centering
\includegraphics[scale = 1]{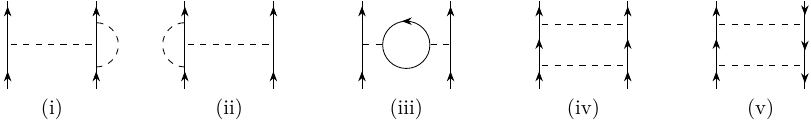}
\\
\vspace{-0.05cm}
\caption{One-loop corrections to the fermion-fermion interactions (the solid and dash lines indicate the fermionic propagator and the fermion-fermion interaction, respectively).}
\label{fig11}
\end{figure*}
\end{widetext}


\end{CJK*}

\end{document}